\documentclass[reprint,tikz,prb,twocolumn,superscriptaddress]{revtex4-1}
\AtBeginDocument{
\heavyrulewidth=.08em
\lightrulewidth=.05em
\cmidrulewidth=.03em
\belowrulesep=.65ex
\belowbottomsep=0pt
\aboverulesep=.4ex
\abovetopsep=0pt
\cmidrulesep=\doublerulesep
\cmidrulekern=.5em
\defaultaddspace=.5em
}

\usepackage{amsmath,amssymb,graphicx}
\usepackage{physics}
\usepackage{float}
\usepackage[usenames,dvipsnames]{xcolor}
\usepackage{tikz,amsthm,comment}
\usepackage{booktabs}
\usepackage[export]{adjustbox}
\usepackage[section]{placeins}
\usepackage[caption=false]{subfig}
\usepackage[percent]{overpic}
\usepackage{lipsum}

\bibpunct{[}{]}{,}{n}{}{}
\usepackage[hidelinks]{hyperref} 
\hypersetup{colorlinks=true,citecolor=Red,linkcolor=Blue,urlcolor=Blue}

\definecolor{ForestGreen}{HTML}{668000}
\definecolor{red1}{HTML}{FF4136}
\definecolor{green1}{HTML}{00802b}

\newcommand{\<}{\langle}
\renewcommand{\>}{\rangle}

\newcommand{\todo}[1]{\textcolor{red}{TODO: #1}} 

\graphicspath{{./OldFigures/}{./NiravFigures/}{./ShiFigures/}}

\begin{document}
%
\title{Magnetic phase transitions in quantum spin-orbital liquids}

\author{Shi Feng}
\affiliation{Department of Physics, The Ohio State University, Columbus, Ohio 43210, USA}
\author{Niravkumar D. Patel}
\affiliation{Department of Physics, The Ohio State University, Columbus, Ohio 43210, USA}
\author{Panjin Kim}
\affiliation{National Security Research Institute, Daejeon 34044, Korea}
\author{Jung Hoon Han}
\affiliation{Department of Physics, Sungkyunkwan University, Suwon 16419, Korea}
\author{Nandini Trivedi}
\affiliation{Department of Physics, The Ohio State University, Columbus, Ohio 43210, USA}

\begin{abstract}



We investigate the spin and orbital correlations of a superexchange model with spin $S=1$ and orbital $L=1$ relevant for $5d^4$ transition metal Mott insulators~\cite{trivedi15}, using exact diagonalization and density matrix renormalization group (DMRG). For spin-orbit coupling $\lambda=0$, the orbitals are in an entangled state that is decoupled from the spins. 
We find two phases with increasing $\lambda$: (I) the S2 phase with two peaks in the structure factor for $\lambda\le\lambda_{c1}\approx 0.34 J$ where $J$ is the ferromagnetic exchange; and, (II) the $S1$ phase for $\lambda_{c1}<\lambda\le\lambda_{c2}\approx 1.2 J$ with emergent antiferromagnetic correlations. Both S1 and S2 phases are shown to exhibit power law correlations, indicative of a gapless spectrum. Upon increasing $\lambda > \lambda_{c2}$ leads to a product state of local spin-orbital singlets that exhibit exponential decay of correlations, indicative of a gapped phase. We obtain insights into the phases from the 
well-known Uimin-Lai-Sutherland (ULS) model in an external field that provides an approximate description of our model within mean field theory.
\end{abstract}
\date{\today}
\pacs{??}

\date{\today}
\pacs{pacs}
\maketitle

\section{Introduction} \label{sec:intro}

Spin-orbit interaction (SOI) and its involved emergent phenomena have been one of the central themes in condensed matter for more than a decade.
In observations of the quantum spin Hall effects~\cite{spinhall04,spinhall05,qsh07}, the discovery of three-dimensional topological insulators~\cite{3d_ti}, and the discovery of quantum spin liquids~\cite{Khaliullin1}, SOI plays a crucial role.
Furthermore in Mott insulators where electron-electron interactions have a substantial effect, SOI is responsible for the new kinds of spin-orbital excitations in heavy transition metal oxides~\cite{kim08,kim09}.

While SOI has opened new directions for the field of Mott physics, much of the focus has been dedicated to understanding of the $d^5$ valence  configuration~\cite{balents10}; relatively less attention has been given to other valence configurations. 
As different fillings lead to different ground states and correspondingly different low energy excitations, it is natural to ask how SOI affects the existing low energy theory of the electronic configurations other than $d^5$.
With this perspective, $d^1$, $d^2$, and $d^3$ Mott insulators have attracted some interest~\cite{balents10-d1,balents11-d2,trivedi13}, while $d^4$ electronic configuration has been put aside until recently~\cite{trivedi15,Trivedi1_d4,Nitin1,Nitin2}. This is mainly due to the expectation that $d^4$ materials remain nonmagnetic in both strong SOI and Hund's coupling limits~\cite{balents11-d2}.
However, a recent study of $d^4$ configurations proposed that a magnetic phase transition is possible in a realistic parameter regime~\cite{trivedi15,Trivedi1_d4}. 
In fact, Ca$_2$RuO$_4$ was shown to have finite 
local moments~\cite{CaRuO_1,CaRuO_2,CaRuO_3}, while 
experiments on double perovskite iridates ~\cite{PeroSkite_1,PeroSkite_2,PeroSkite_3,BaYIrO_1,BaYIrO_2,BaYIrO_3,BaYIrO_4}, honeycomb ruthenates \cite{Ruthnate_1,Ruthnate_2,Ruthnate_3} have revealed magnetism as a key player.


Motivated by our earlier work, we examine the low-energy effective Hamiltonian describing one-dimensional $d^4$ transition metal oxides, proposed by Ref.~\cite{trivedi15}. We remark that Ref.~\cite{trivedi15} studied the full multi-orbital Hubbard model in search of the unusual magnetic phase transition mediated by SOI in systems on $2$-sites with $d^4$ configuration. 
Here, we study the effective low-energy spin-orbital Hamiltonian on large systems using exact diagonalization (ED) and density matrix renormalization group (DMRG). Unlike the $2$-site study that only show one phase transition, in this article we demonstrate that there are two distinct phase transitions with increasing SOI. In this article, (1) we identify each phase in the phase diagram, (2) provide the explicit ground-state in extreme limits of SOI, (3) demonstrate spin-orbital separation in the small SOI limit, and (4) demonstrate that our low energy effective model of $d^4$ configuration is equivalent to a well known Uimin-Lai-Sutherland (ULS) model ~\cite{ULS1,ULS2,ULS3} with a small external field. 

This work is organized as follows. Section II 
introduces the model and the methods. Section
III presents the main results. Section IV establishes the 
correspondence between the magnetic Hamiltonian describing the $d^4$ configuration 
and the ULS model in an external field.
Section V provides the final conclusions. 

\section{Model and Methods} \label{sec:MM}

The superexchange Hamiltonian we investigate is obtained from a microscopic electronic Hamiltonian, as shown previously~\cite{trivedi15}, and is given by:
%
\begin{equation} \label{eq:hamiltonian1}
    \begin{split}
      H_{d4} = &-J_{FM} \sum_{\< ij \> }(\mathbf{S}_i \cdot \mathbf{S}_j ) \mathcal{P}(\mathbf{L}_i + \mathbf{L}_j = 1) \\ 
      &+ \lambda \sum_i  \mathbf{L}_i \cdot \mathbf{S}_i,
  \end{split}
\end{equation}
%

%
\begin{figure}[t]
  \centering
    \includegraphics[trim={0.0cm 0 -0.0cm 0},width=0.45\textwidth]{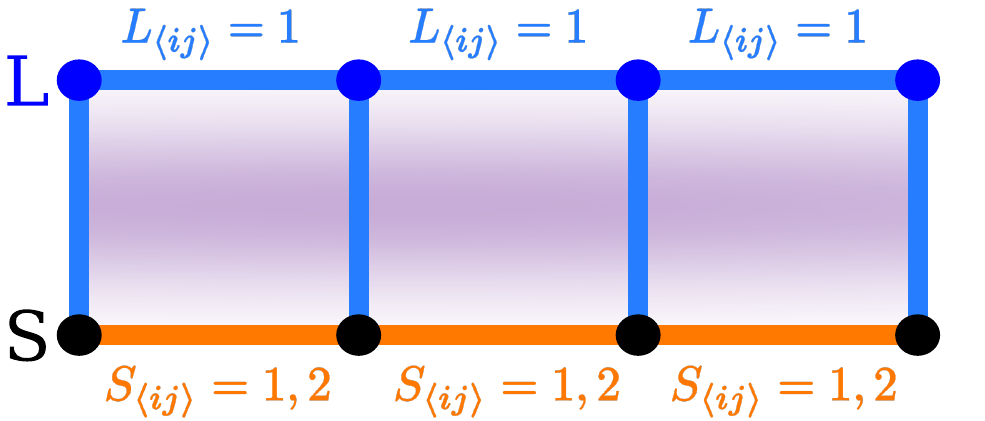}
    \caption{
    Schematic representation of our model (Eq.~\ref{eq:hamiltonian1}) on a chain with $4$-sites. Each site has a spin $S$ (black circle) and orbital $L$ (blue circle) degrees of freedom. The blue (red) bonds represents effective anti-ferromagnetic (ferromagnetic) interactions. In addition, nearest neighbor $L-L$ bonds are projected into local bond $L_{\<i j\>} = 1$ while the spin bonds $S-S$ are projected into $S_{\<i j\>} = 1$ or $2$. The vertical blue bonds represents AFM spin-orbit coupling. Finally, the purple shade represents interactions between the $2$ spins and $2$ orbitals resulting from the projector in (Eq.~\ref{eq:hamiltonian1}).
    }
\vspace{-0.35cm}
\label{fig:schematic}
\end{figure}
%
\noindent
where $\mathbf{S}_i$ and $\mathbf{L}_i$ are local spin $S=1$ and orbital angular momentum $L=1$ operators at the site $i$ with $\< i j\>$ representing the nearest-neighbor sites. The coupling constants $J_{FM}$ and $\lambda$ represent ferromagnetic exchange and spin-orbit interactions. The projection operator in the first term, $\mathcal{P}(\mathbf{L}_i + \mathbf{L}_j = 1)=- \frac{1}{8} \mathbf{L}_{\<ij\>}^2 (\mathbf{L}_{\<ij\>}^2 - 6)$, is defined on a bond connecting orbital sectors of two adjacent sites. For a two site problem, the total orbital angular momentum can be $L_T = 0,1$ or $2$. Therefore, the projector $\mathcal{P}(\mathbf{L}_i + \mathbf{L}_j = 1) = 0, 1, 0$ for $L_T = 0,1,2$ respectively. 
Considering the overall minus sign in the first term of the Hamiltonian ($-J_{FM}$), this  projector makes the $L_{\<i j\>} = 0$ and $2$ quantum sectors energetically unfavorable (projects out) on the two-site bond, while preferring $L_{\<i j\>} = 1$ angular momentum on the bond. Figure~\ref{fig:schematic} shows a schematic representation of the $H_{d4}$ model where intra-orbital (blue sites) bonds are projected into the Hilbert space of $L_{\<ij\>}=1$ on each bond. Expanding the projector reveals the explicit form of the Hamiltonian:  
%
\begin{equation}\label{eq:hamiltonian}
    \begin{split}
          H_{d4} = &\frac{J_{FM}}{2} \sum_{\< ij \> }(\mathbf{S}_i \cdot \mathbf{S}_j ) \left( (\mathbf{L}_i \cdot \mathbf{L}_j)^2 + \mathbf{L}_i \cdot \mathbf{L}_j - 2 \right) \\
     & + \lambda \sum_i \mathbf{L}_i \cdot \mathbf{S}_i, 
    \end{split}
\end{equation}
%
where the ferromagnetic spin exchange is illustrated by intra-spin red bonds in Fig.~\ref{fig:schematic}. 
We solve this model using a combination of exact diagonalization (ED) and density matrix renormalization group (DMRG) with $S,L = 1$. ED is used for $4$-site chain calculations with a $3^{8}$ dimensional Hilbert space under
periodic boundary conditions, while DMRG is used to solve the $64$-site chain with open boundary conditions.
In order to lift a large ground-state degeneracy, we apply a small a pinning field $10^{-5}$ at the edge of the chain. This avoids random linear combinations of degenerate states and allows calculations of numerically consistent results at all values of $\lambda$. 
All DMRG calculations were performed using the {ITensor} Library~\cite{ITensor} with maximum number of kept states $m=800$ with a fixed truncation error $10^{-7}$. 

\subsection*{Operators and Observables}
In this section, we describe the observables whose behavior is presented in the Results section. The gap in the energy spectrum is defined by
%
\begin{equation} \label{eq:dE}
    \begin{split}
        \Delta E = E_m - E_0, 
    \end{split}
\end{equation}
%
is the difference in energy $E_m$ of the $m^{th}$ state from that of the ground state $E_0$.
We use the operator $O$ to represent spin $S$, orbital $L$ or total $J$ angular momentum operators, and subscripts $i$ and $j$ are label the sites with a total of $N$ sites in the chain. The total quantum numbers $O_T$ of the full system are defined using  
%
\begin{equation} \label{eq:QuantNumb}
    \begin{split}
        O_T^2 &= O_T (O_T + 1) = \sum_{i,j} \< \mathbf{O}_i \cdot \mathbf{O}_j \>. 
    \end{split}
\end{equation}
%
To make our analysis easier, we use this definition even when $O_T$ is not quantlized. 
The total magnetization related to the $O$ angular momentum is defined as 
%
\begin{equation} \label{eq:Mag}
    \begin{split}
        M_{O} &= \frac{1}{N} |\sum_{i} \< O^z_i \>|,  
    \end{split}
\end{equation}
%
where only the $z$ projection of $O$ angular momentum is used. We also present calculations of connected and un-connected 
correlations in the main text. The connected correlator 
is defined as 
%
\begin{equation} \label{eq:ConnCorr}
    \begin{split}
        \< \mathbf{\delta O}_{i} \cdot \mathbf{\delta O}_{j} \> &= \< \mathbf{O}_i \cdot \mathbf{O}_j \> - \< \mathbf{O}_i \> \cdot \< \mathbf{O}_j \> 
    \end{split}
\end{equation}
where the ground-state expectation value $\< \mathbf{O}_i \> \cdot \< \mathbf{O}_j \>$ are subtracted from the correlations $\< \mathbf{O}_i \cdot \mathbf{O}_j \>$. We present the real-space correlations $C_O(R)$ to study the 
decay of correlations and $G_O(R)$ to study the real-space alignment of $O$ with respect to a site of reference $i_{r}$. 
%
\begin{equation} \label{eq:RealSpCorr}
    \begin{split}
        C_O(R) &= \frac{1}{N_R} |\sum_{i} \< \mathbf{\delta O}_{i} \cdot \mathbf{\delta O}_{i+R} \>|, \\
        G_O(R) &= \< \mathbf{\delta O}_{i_{r}} \cdot \mathbf{\delta O}_{i_{r}+R} \>, 
    \end{split}
\end{equation}
%
where $N_R$ refers to the number of nearest-neighbor pairs at a distance $R$ relative to a chosen reference site. 
We also calculate the momentum space correlations
%
\begin{equation} \label{eq:Ok}
    \begin{split}
        O(k) &= \frac{1}{N^2} \sum_{i,j} e^{ik(r_i - r_j)} \< \mathbf{\delta O}_{i} \cdot \mathbf{\delta O}_{j} \>, 
    \end{split}
\end{equation}
%
using the Fourier transform of $\< \mathbf{\delta O}_{i} \cdot \mathbf{\delta O}_{j} \>$ in order to elucidate the underlying quasi-long-range order. The $r_i$ and $r_j$ refers to the real-space coordinates of sites $i$ and $j$ and the $k$ represents the crystal momentum.

\section{Results}
%
\begin{figure*}[t]
  \centering
    \includegraphics[trim={0cm 0cm -0.0cm 0cm},width=0.98\textwidth]{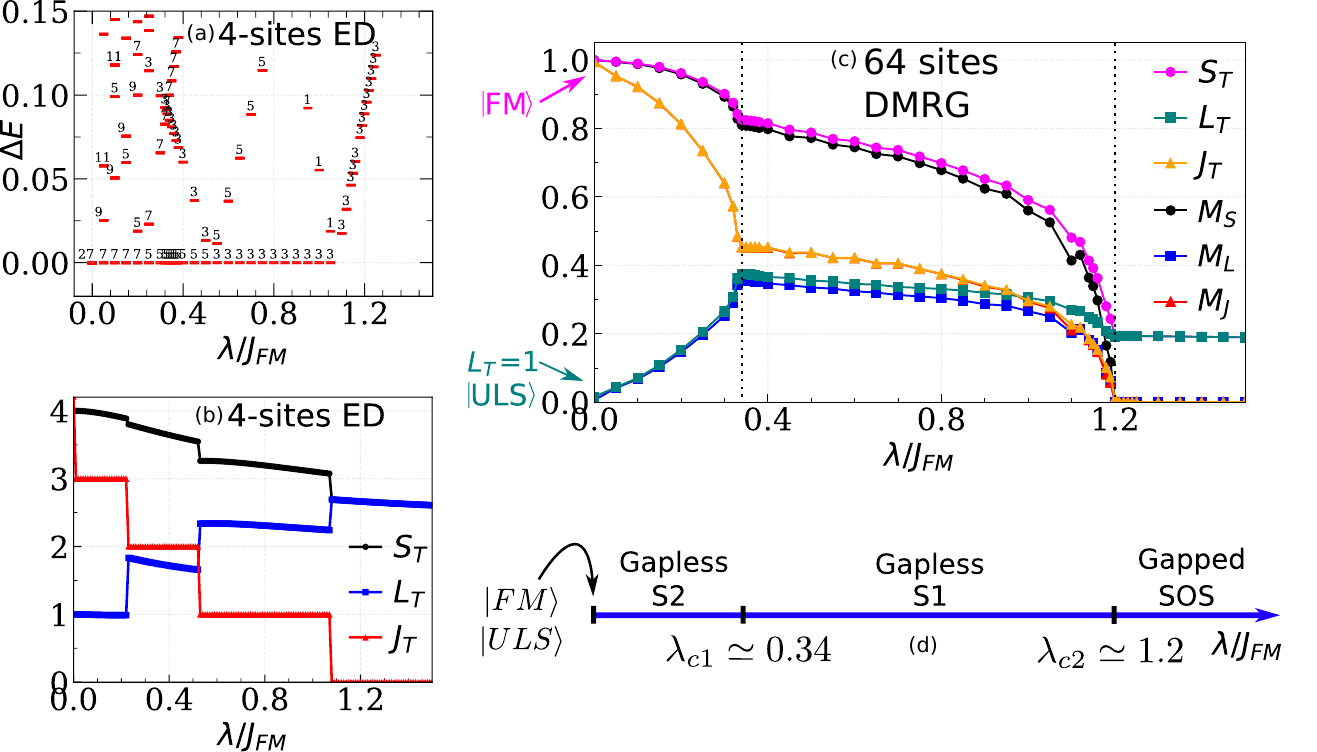} 
    \subfloat{\label{fig:2a}}
    \subfloat{\label{fig:2b}}
    \subfloat{\label{fig:2c}}
    \subfloat{\label{fig:2d}}
    \subfloat{\label{fig:2e}}
    \caption{
    (a) Energy gap spectrum (Eq.~\ref{eq:dE}) and the (b) spin $(S_T)$, orbital $(L_T)$ and total $(J_T)$ angular momentum (Eq.~\ref{eq:QuantNumb}) as a function of spin-orbit coupling ($\lambda$) for $4$ sites chain. Results of (a-b) are obtained using exact diagonalization. 
    The numbers in the spectrum represents degeneracy of the corresponding state (red-lines).
    (c) The total spin, orbital and $J$ magnetization (Eq.~\ref{eq:Mag}) and $S_T$, $L_T$, $J_T$ per site as a function of $\lambda$. Results of (c-d) are obtained using DMRG calculations on a $64$-site chain. 
    (d) phase diagram representing four different phases with two critical fields $\lambda_{c1}$ and $\lambda_{c2}$. 
    }
\vspace{-0.2cm} \label{fig:2}
\end{figure*}
%
We begin by describing our results in the two extreme limits 
of the SOI: $\lambda =0$ and $\lambda \rightarrow \infty$. 

\noindent
{\bf \boldmath$\lambda = 0$ limit ---}
Figure~\ref{fig:2a} and \ref{fig:2b} show the energy spectrum and quantum numbers
as a function of spin-orbit coupling $\lambda$, obtained using ED on a $4$ sites periodic chain.
We find a large $27$-fold degeneracy in the ground-state (g.s.) for $\lambda = 0$.
To understand this degeneracy, we show measures of total angular momentums in the 
Figure~\ref{fig:2b}. The $\lambda = 0$ Hamiltonian commutes with the
total $S^2$ and orbital $L^2$ angular momentum, defining $S_T$ and $L_T$ as `good' quantum numbers. The $\lambda = 0$ g.s. has $S_T = 4$ (maximum for $4$-sites chain) quantum number with $L_T = 1$, leading to $2S_T + 1 = 9$ fold spin degeneracy and $2L_T + 1 = 3$ fold orbital degeneracy (Fig.~\ref{fig:2b}). Therefore, the total g.s. degeneracy is $27$ as shown in Figure~\ref{fig:2a}.
This implies that the spin and orbital degrees of 
freedom are approximately decoupled for $\lambda=0$ phase. This is surprising because SOI is still present through the terms $(\mathbf{S}_i \cdot \mathbf{S}_j ) \left( (\mathbf{L}_i \cdot \mathbf{L}_j)^2 + \mathbf{L}_i \cdot \mathbf{L}_j \right)$ even at $\lambda=0$.   
Overall, the g.s. of $\lambda =0$ can be expressed as 
\begin{equation} \label{eq:ULS}
|\psi_0 \>_{\lambda = 0} \simeq|S_T = \text{N}\> \otimes |L_T = 1\>
\end{equation}
where $|S_T = \text{N}\>$ represents a ferromagnet (Figs.~\ref{fig:2b} and \ref{fig:2d}). 
{\it This wave-function factorization is a signature of the spin-orbital separation in a strongly-interacting system}. 
Considering this factorized g.s., if we ignore the spin terms in the Hamiltonian, we obtain the effective orbital model as the well-known Uimin-Lai-Sutherland (ULS) Hamiltonian. 
%
\begin{equation} \label{eq:ULS}
H_{ULS} = \sum_{\< ij \> }\left( (\mathbf{L}_i \cdot \mathbf{L}_j)^2 + \mathbf{L}_i \cdot \mathbf{L}_j - 2 \right)
\end{equation}
%
The ULS model is exactly solvable and its ground-state ($|ULS\>$) is well known ~\cite{ULS1,ULS2,ULS3}.
The ground state of our Hamiltonian 
$H_{d4}$ in the $\lambda=0$ spin-orbital factorized phase can therefore be written as  
%
\begin{equation} \label{eq:FMULS}
|\psi_0 \>_{\lambda = 0} \simeq |FM\> \otimes |ULS\>. 
\end{equation}
%

We have extended our results of the $4$-sites chain by performing large-scale DMRG simulations on $64$ sites chains (Fig.~\ref{fig:2c} and \ref{fig:2d}). We find that $L_T = 1$ and $S_T = N$ for the $\lambda = 0$ g.s., in agreement ferromagnetic ULS-orbital state understood from the ED results. In addition, the large  magnetization 
$M_S$ clearly indicates a spin ferromagnetic g.s. for $\lambda = 0$ (Fig.~\ref{fig:2d}). We remind the readers here that by using a pinning field in the DMRG calculations, we `pick' only the $M_S=N$ state from the $2N+1$-fold degenerate FM g.s. state.

\noindent
{\bf \boldmath$\lambda \rightarrow \infty$ limit ---}
Upon increasing $\lambda$, the large g.s. degeneracy decreases, eventually resulting in a unique g.s. for $\lambda \gtrapprox 1.1$ (Fig.~\ref{fig:2a}). For $\lambda \rightarrow \infty$, only the on-site spin-orbit interactions dominate whereas the inter-site interactions become negligible. Therefore, the g.s. is simply composed of on-site anti-aligned spin-orbital singlet $|J = 0 \>_i$. The full g.s. is represented as a product-state:   
%
\begin{equation} \label{eq:product}
|\psi_0\>_{\lambda \rightarrow \infty} = \prod_{i=1}^{N} |J = 0 \>_i, 
\end{equation}
%
and we refer to this phase as the non-degenerate spin-orbital singlet (SOS) phase. 
In this state, $J_T = 0$ while 
$L_T^2 = L_T(L_T+1) = S_T^2 = S_T(S_T+1) = 2N$. This is due to the negligibly weak coupling between adjacent orbitals at large $\lambda$, at which only the diagonal terms in Eq.~\ref{eq:QuantNumb} survive:
\begin{equation}
\<\textbf{S}_T^2\>_{\lambda \rightarrow \infty} =  \sum_{i}^N \expval{\textbf{S}_i^2 } =\sum_{i}^N S_i(S_i+1)  
\end{equation}
Hence in SOS phase $S_T = L_T = 10.83$ for $64$ sites chain, in agreement with results of Fig.~\ref{fig:2d}. The same is true for $4$-sites ED results with $S_T = L_T = 2.37$ (Fig.~\ref{fig:2b}). Additionally, the energy spectrum is gapped within the SOS phase with a gap value that increases linearly with $\lambda$. The first excited state is simply $|J=1\>$ on-site spin-orbital triplet with 3-fold degeneracy (Fig.~\ref{fig:2a}).  
\begin{figure*}[t]
  \centering
    \includegraphics[trim={0cm 0cm -0.0cm 0cm},width=0.98\textwidth]{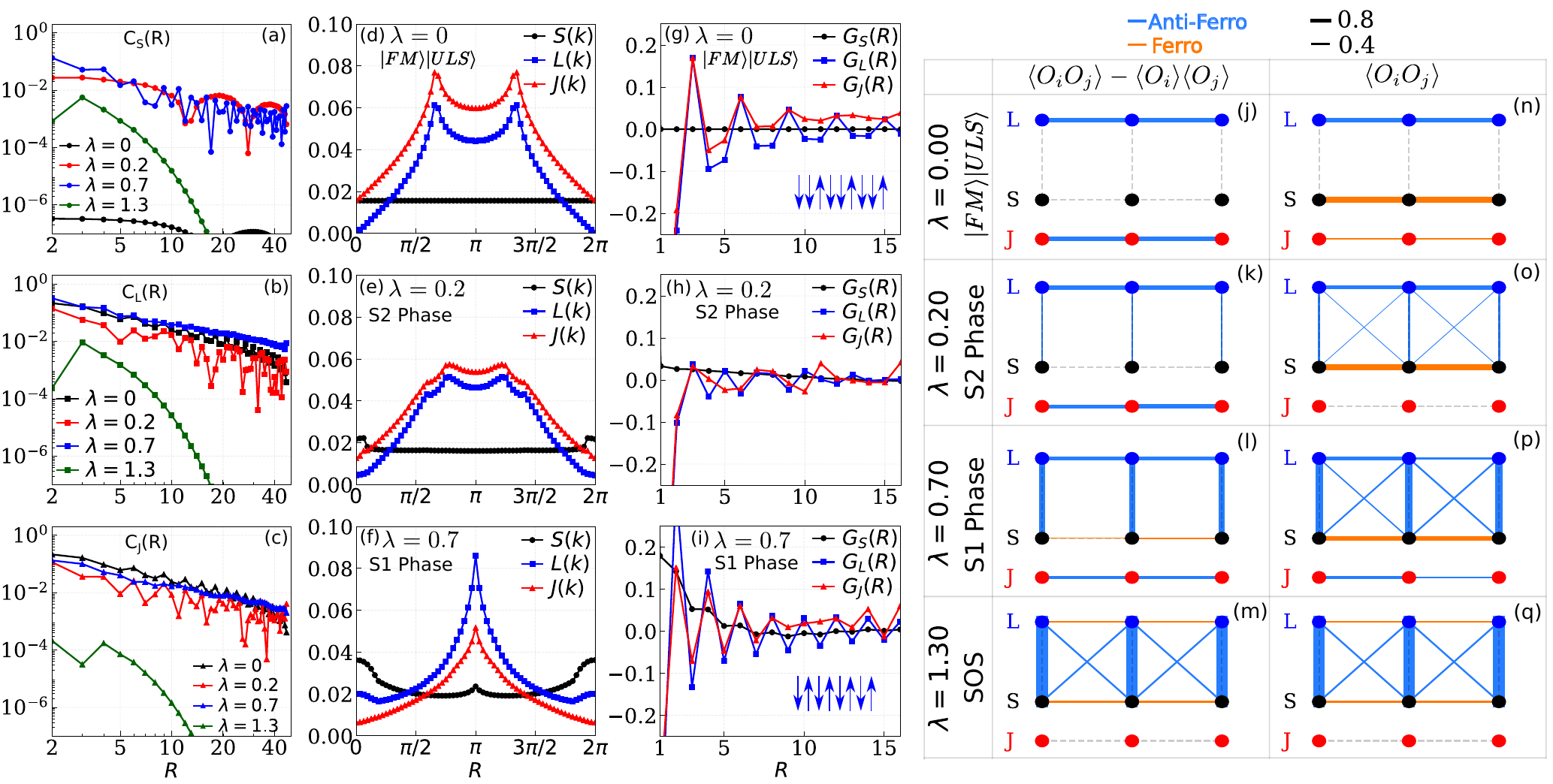}
    \subfloat{\label{fig:3a}} \subfloat{\label{fig:3b}} \subfloat{\label{fig:3c}} \subfloat{\label{fig:3d}} \subfloat{\label{fig:3e}}
    \subfloat{\label{fig:3f}} \subfloat{\label{fig:3g}} \subfloat{\label{fig:3h}} \subfloat{\label{fig:3i}} \subfloat{\label{fig:3j}}
    \subfloat{\label{fig:3k}} \subfloat{\label{fig:3l}} \subfloat{\label{fig:3m}} \subfloat{\label{fig:3n}} \subfloat{\label{fig:3o}}
    \caption{
    Decay of the (a) spin, (b) orbital and (c) total $J$ correlations (Eq.~\ref{eq:RealSpCorr}) for various values of $\lambda$. (d,e,f) momentum-space (Eq.~\ref{eq:Ok}) and (g,h,i) real-space correlations (Eq.~\ref{eq:RealSpCorr}) for $\lambda = 0$, $\lambda = 0.2$ S1 phase, and $\lambda = 0.7$ S2 phase (top to bottom panel respectively). 
    (j-q) On-site and the nearest neighbor real-space bulk (around the central sites) correlations for $\lambda =0$, S2 phase, S1 phase and the SOS phase (top to bottom panel) where left column represents the connected correlator (Eq.~\ref{eq:ConnCorr}) and right column represents the total correlations. The blue and orange line represents ferromagnetic (alignment) or anti-ferromagnetic (anti-alignment) correlations, while the line thickness represents the magnitude of the correlations. In each panel, orbital, spin and total $J$ sites are filled circles with colors blue, black and red (respectively). All Results are obtained using $64$ sites DMRG simulation. 
    }
\vspace{-0.2cm}
\label{fig:3}
\end{figure*}
%
\vspace{0.2cm}

\noindent
{\bf \boldmath$\lambda \ne 0$ ---}
The spin and orbital angular moments are not quantized as $S^2$ and $L^2$ do not commute with the Hamiltonian, nevertheless the total $J^2$ commutes with $H_{d4}$ for all $\lambda$ values. 
$J_T$ decreases with increasing $\lambda$ (Fig.~\ref{fig:2b}), and each quantized value of $J_T$ leads to $2J_T+1$ fold degeneracy in the g.s. (Fig.~\ref{fig:2a}). Remarkably, as shown in the 4-site ED results, an energy level crossing occurs near the g.s. at $\lambda=0.5$. The level crossing in the g.s. energy at $\lambda=0.5$ in the $\Delta E$ is indicative of an unexpected intermediate phase transition, which is rounded due to the small number of sites. This intermediate phase is revealed more clearly through kinks in the magnetization and total quantum numbers obtained from DMRG calculations (Fig.~\ref{fig:2c} and \ref{fig:2d}). We emphasize that our discovery of this intermediate phase is new and has not been reported in the previous study investigating a $2$-site model ~\cite{trivedi15}. 

We identify two critical points in Figure~\ref{fig:2e} phase diagram at $\lambda_{c1} \simeq 0.34$ and $\lambda_{c2} \simeq 1.20$. We label $0.0 < \lambda < 0.34$ phase as the `S2' phase and $0.34 < \lambda < 1.20$ as the `S1' phase (Fig.~\ref{fig:2a}), following the terminology used in Ref.~\cite{littlewood98}. Overall, we identify $4$ different phases that we further explore using g.s. spin, orbital and total angular momentum ($J$) correlations. 

Figure~\ref{fig:3} summarizes results of all two-point correlations in the different 
phases of the phase diagram. Figures~\ref{fig:3}(a-c) shows the decay of 
the spin $S$, orbital $L$ and $J$ connected correlations in each phase. All correlations in the intermediate S1 and S2 phases have a power-law decay, while, all correlations decay exponentially for $\lambda > 1.3$ in the SOS phase, as expected for a product state Eq.~\ref{eq:product}). Finally, for $\lambda=0$, the orbital and $J$ correlations decay as a power-law decay, but the spin correlations show an exponential decay. 
We remark that exponential decay of the spin correlations is likely a consequence of our numerical method that `picked' only the $M_S = N$ state (due to the pinning field in the DMRG calculation) from the $2N+1$ spin-degenerate manifold of the $|FM\>$ state. 
In fact, $C_s(R) < 10^{-6}$ for $\lambda=0$ and therefore should be considered to be zero within our numerical accuracy.  We remark that upon considering the full degenerate FM sector, the spin correlations for $\lambda=0$ may have a power-law decay, but is difficult to capture numerically because of the large degeneracy. Regardless of the spin decay, we show that the orbital and therefore the total $J$ correlations decay as a power-law. This implies an overall gapless energy spectrum for $\lambda=0$, S1 and S2 phases as summarized in the phase diagram (Fig.~\ref{fig:2e}). 

Figures~\ref{fig:3}(d-f) show the spin, orbital and $J$ connected correlations in momentum space. At $\lambda = 0$, we find a peak at the incommensurate crystal momentum $k \simeq 0.6 \pi$ in $L(k)$. Note that this peak is expected because the $\lambda=0$ g.s. is factorized such that the orbital angular momentum part of the g.s. is effectively represented as the g.s. of the ULS Hamiltonian (Eq.~\ref{eq:FMULS}). A peak at the $k \simeq 0.6\pi$ is known to be present in the ULS g.s. Figure~\ref{fig:3g} shows the corresponding spin, orbital and $J$ correlations in the real-space. The fluctuating blue 
curve allow us to write a `classical' state with a series of repeated up-down-down orbital $L$ patterns, represented by the blue arrows in Figure~\ref{fig:3g}. 
The Figures~\ref{fig:3}j/n show a real-space cartoon of correlations between the nearest-neighbor $L$ and $S$ with anti-ferro blue bonds and ferro red bonds. For $\lambda=0$ spin angular momentum, we expect to see a peak at $k=0$ for the spin $S(k)$ because the spin part of the factorized g.s. is described by the $S_T = N$ ferromagnet. However, due to capturing only $M_S = N$ product-state of the full $2N+1$ fold degenerate spin sector, we do not find a peak in the $S(k)$ even though we have already established that $\lambda=0$ state to be a spin ferromagnet. Figures~\ref{fig:3}j/n further corroborate our explanation where the connected correlations between spins are zero, yet the correlations $\< \mathbf{S}_i \cdot \mathbf{S}_j \>$ clearly show ferromagnetic orange bonds between the nearest-neighbor spins. Additionally, all correlations between $S$ and $L$ are zero, consistent with the g.s. $|FM\> \otimes |ULS\>$ ansatz.   

Upon increasing $0<\lambda<0.34$ into the S2 phase leads to bifurcation of the $L(k=0.60\pi)$ peak into two features that separate with increasing $\lambda$. This multi-peak structure leads to interference between different momenta, leading to real-space structure that is difficult to decipher (Fig.~\ref{fig:3h}). The corresponding real-space nearest-neighbor correlations cartoon show weak correlations between the spin $S$ and $L$ angular momentum (Fig.~\ref{fig:3}k/o). 

Beyond $\lambda_{c2}$, $L(k)$ shows a peak at $k=\pi$ in the S1 phase ($\lambda=0.7$) representing a anti-ferro orbital alignment as shown by the short period fluctuating blue curve in real space Figures~\ref{fig:3i}. However, we remark that the S1 phase is not a usual N\'eel type anti-ferro orbital order because $L_T \ne 0$ in this case, unlike the singlet N\'eel type AFM g.s.of the Heisenberg model (Fig.~\ref{fig:2d}). Additionally, the coupling between the spins and orbital is robust as represented by blue vertical on-site $L-S$ bonds in Figures~\ref{fig:3}l/p. Finally for $\lambda = 1.3$ in the SOS phase, all nearest-neighbors correlations are significantly suppressed while on-site $S-L$ bonds remains robust, consistent with the product-state ansatz (Eq.~\ref{eq:product}). For all phases, we also show the total $J$ correlations qualitatively behaves as the $L$ correlations(Fig.~\ref{fig:3}). 

\section{Discussion and ULSZ Model correspondence}
Over the range of $\lambda \in [0, \infty]$, there exist four different phases which we call $|FM\>|ULS\>$, S2, S1 and spin-orbital singlet (SOS) states. An uncanny resemblance of the phase diagram of our model (Eq.~\ref{eq:hamiltonian}) to a seemingly unrelated model studied several decades ago~ \cite{littlewood98} lead us realize the following. For the spin-orbital coupling $\lambda=0$, we have demonstrated that the g.s. factorize into FM spin state and `ULS' orbital state (Eq.~\ref{eq:FMULS}). 
Realizing this, we can also factorize our Hamiltonian into a spin and orbital part, similar to a mean-field approximation, 
%
\begin{equation} \label{eq:eff_hamiltonian}
    \begin{split}
        H^{eff} = &\frac{J_{FM} M_S^2}{2} \sum_{\langle ij\rangle } \Bigl( (\mathbf{L}_i \cdot \mathbf{L}_j)^2 + \mathbf{L}_i \cdot \mathbf{L}_j -2 \Bigr) 
        \\ & + \lambda M_S \sum_{i} L_i^z  
    \end{split}
\end{equation}
%
where we assume $M_s^2 \simeq \sum_{\<ij\>} \< \mathbf{S}_i \cdot \mathbf{S}_j \>$ and treat the spin-orbit coupling in the Ising limit with $M_S = \sum_i |\<S_i^z\>|$.
In summary, the similarity with Ref.~\cite{littlewood98} lead us to expect that our model (Eq.~\ref{eq:hamiltonian1}) can be well approximated by only the orbital term in the $H_{d4}$ Hamiltonian with a Zeeman field $L^z$ (Eq.~\ref{eq:eff_hamiltonian}). The first term in $H^{eff}$ is exactly the ULS Hamiltonian (Eq.~\ref{eq:ULS}) up to a constant. Therefore the effective Hamiltonian can be interpreted as the ULS Hamiltonian with an additional Zeeman field, which we dub the ULSZ Hamiltonian, given by:
%
\begin{align}\label{eq:ULSZ_hamiltonian}
  H_{\rm ULSZ} =& \sum_{\langle ij\rangle } \Bigl( (\mathbf{L_i} \cdot \mathbf{L_j})^2 + \mathbf{L_i} \cdot \mathbf{L_j} \Bigr) + h \sum_i L_i^z ,
\end{align}
%
where $\mathbf{L}_i$ are the spin-1 Pauli operators at site $i$, and $h$ is the strength of an external magnetic field applied along the $z$-direction.
%
\begin{figure}[t]
  \centering
    \includegraphics[trim={0.1cm 0 -0.1cm 0},width=0.5\textwidth]{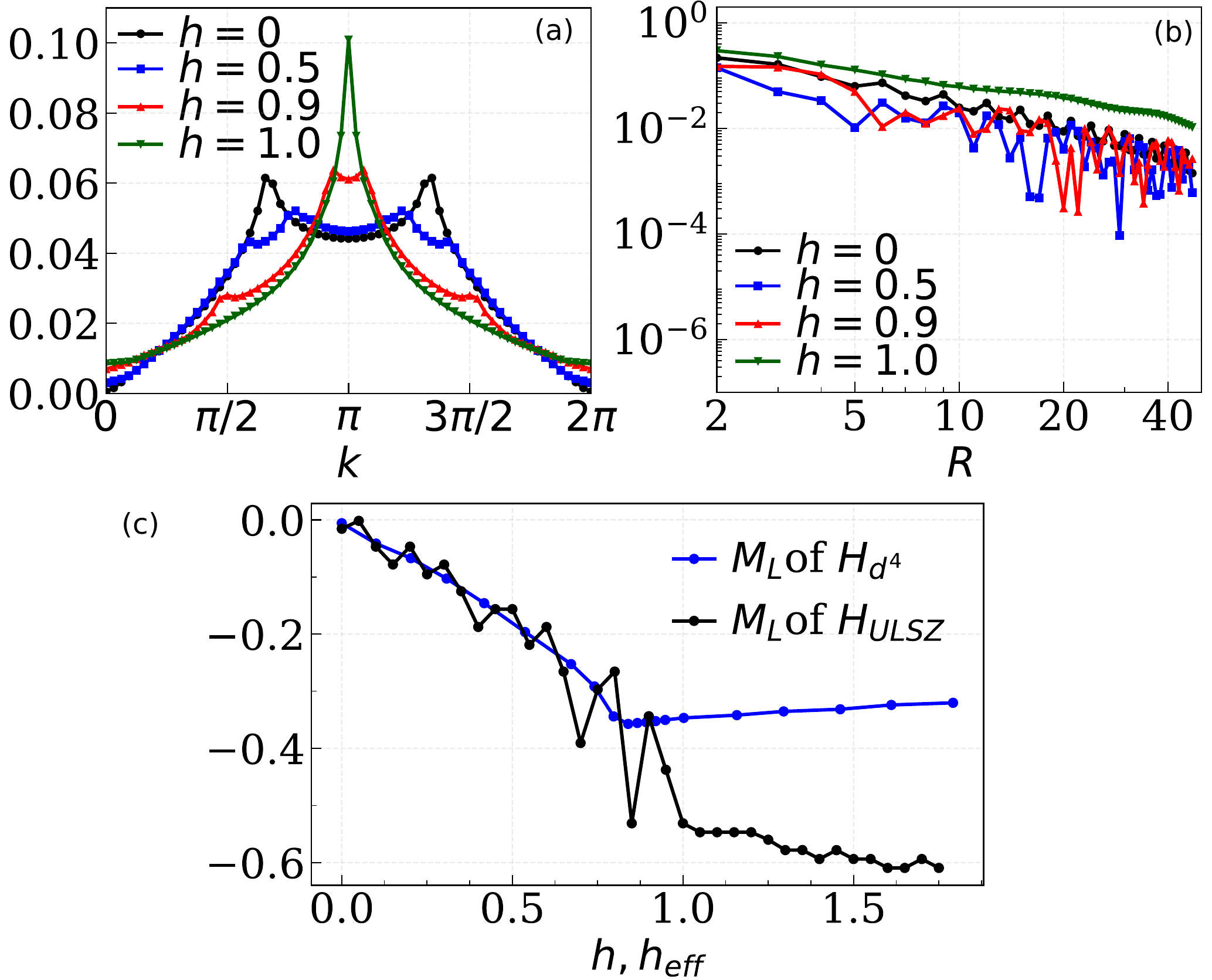}
        \subfloat{\label{fig:4a}} \subfloat{\label{fig:4b}} \subfloat{\label{fig:4c}}
    \caption{
    (a) momentum-space and (b) real-space correlations at $h = 0, 0.5, 0.9, 1.0$ (respectively). (c) $\langle \bar{S^z} \rangle$-$h$ curve of the ULSZ chain of length $L=64$ (black) and $\langle \bar{L^z} \rangle$-$h^{\rm eff}$ curve (blue) of $d_4$ chain in S2 and part of S1 phase. Both curves almost overlap for  $0 < h^{\rm eff} < 0.83$. The $y$-axis is in units of $\hbar$. The inserted graph is the zoom-out of ULSZ $\langle \bar{S^z} \rangle$ on a wider range of $h$, which shows a transition to an intermediate phase at $h = 0.9$ , and transition to a fully polarized state at $h = 4.0$. Results are obtained using 64 sites DMRG with OBC. \\}
\vspace{-0.2cm}
\label{fig:ULSZ}
\end{figure}
%

Figure~\ref{fig:ULSZ} summarizes the results of the ULSZ model. At $h = 0$, in the ground state of the ULS Hamiltonian, the real space correlation function exhibits a power law decay. This is consistent with the well-known fact that the g.s. of ULS model is gapless. The SU(3) symmetry of the ULSZ model at $h=0$ is immediately broken down to U(1)$\times$U(1) at $h>0$. The model remains gapless for $0<h<h_{c1} = 0.941$ in the S2 phase where two gapless modes characterize the low-energy excitations~\cite{littlewood98}. For $h_{c1} < h < h_{c2} = 4$ in the S1 phase, one gapless branch is realized~\cite{littlewood98}. And for $h>h_{c2}$ the system becomes fully spin-polarized phase. 
These results are consistent with the power-law decay of S1 and S2 phase of $H_{ULSZ}$ (Fig.~\ref{fig:4b}) that implies a gapless phase. Moreover, the power-law decay of the $h=0$, S1 and S2 phase are also consistent with the results of the $H_{d4}$ (Eq.~\ref{eq:hamiltonian1}). 

Figure~\ref{fig:4a} shows the momentum-space correlations. The incommensurate orbital ordering is demonstrated by a peak for $k \simeq 0.6\pi$ which bifurcates into two incommensurate peaks with increasing $h$ (S2 phase) followed by one-peak structure at $k=\pi$ for the S1 phase ($0.941 < h < 4$). This is also consistent with the momentum-space correlations of $H_{d4}$ (Figure~\ref{fig:3}d-f).
Overall, the decay correlations and the quasi-long-range order resembles the results of $H_{d4}$, {\it confirming that the low energy spin-orbital $H_{d4}$ model can be approximated with a ULS model with an external field}.


It is necessary to ascertain to what extent is the ULSZ an accurate approximation of the $H_{d4}$ Hamiltonian 
(Eq.~\ref{eq:hamiltonian}). 
Comparing the effective $H^{eff}$ with the $H_{ULSZ}$ (Eq.~\ref{eq:eff_hamiltonian} and \ref{eq:ULSZ_hamiltonian}), it is 
immediately clear that 
%
\begin{align}\label{eq:heff}
  h \leftrightarrow h_{eff} = { 2\lambda \over M_S}.
\end{align}
%
The first critical point in the $H_{d^4}$ model occurs at $\lambda_{c1} = 0.34$, at which $M_S = 0.81$ (Fig.~\ref{fig:2c}). This can be translated into the effective Zeeman field $h = 2\lambda_{c1} / M_S = 0.83$ which qualitatively agrees with the value of the first critical point in ULSZ model at $h_{c1} \simeq 0.9$, as shown in Fig.~\ref{fig:4c}. Moreover, the estimates of $M_L$ of the two models roughly coincide for $h<h_{c1}=0.9$. However, the orbital magnetization $M_L$ becomes distinct for $h>0.9$ in the S1 phase  even though the corresponding orbital correlations in the S1 phase show a peak at $k=\pi$, in agreement with the results of $H_{d4}$ (Fig.~\ref{fig:3f}). 

Finally, the results of $H_{d4}$ and $H_{ULSZ}$ are explicitly different in the large field $h$ phase. In the $H_{ULSZ}$ model, large field leads to a product-state where all the orbitals are aligned along the $L^z$ projection, while $H_{d4}$ model leads to a local spin-orbital singlet product state for $\lambda>1.2$ (Eq.~\ref{eq:product}). 
We have illustrated $H_{d4}$ Hamiltonian is analogous to the $H_{ULSZ}$ Hamiltonian in the low field $h$ limit. In fact, $H_{ULSZ}$ can qualitatively capture the S1 and S2 phases found in the $H_{d4}$ Hamiltonian. 

\section{Summary} \label{sec:summary}
In this paper, we have investigated the low energy magnetic Hamiltonian resulting from an electronic Hamiltonian pertinent for the $5d^4$ electronic configuration of transition metal oxides with competition between superexchange interaction and strong SOI.
For spin-orbit coupling $\lambda=0$, we found a wave-function factorization of two orbital sectors, which is a smoking-gun signature of the emergent spin-orbital separation in a spin-orbital interacting system.
Two phases are found thereafter with increasing $\lambda$: the S2 phase with two peaks in the structure factor for $\lambda\le\lambda_{c1}\approx 0.34 J$ where $J$ is the ferromagnetic exchange, and the $S1$ phase $\lambda_{c1}<\lambda\le\lambda_{c2}\approx 1.2 J$ with antiferromagnetic correlations. The $\lambda=0$, S2 and S1 phases are shown to exhibit power law correlations, indicative of a gapless phase. Increasing $\lambda > \lambda_{c2}$ leads to a product state of local spin-orbital singlets that exhibits exponential decay of correlations, indicative of a gapped phase.
By a mean-field like approximation, we demonstrated the correspondence between the $H_{d4}$ at low energy with the well-known Uimin-Lai-Sutherland (ULS) Hamiltonian with an external field (ULSZ). The results of $H_{d4}$ and $H_{ULSZ}$ are explicitly different in the large field $h$ phase. However, in the $H_{ULSZ}$ model, large field leads to a product-state where all the orbitals are aligned along the $L^z$ projection, while $H_{d4}$ model leads to a local spin-orbital singlet product state for $\lambda>1.2$. Hence the approximation breaks down eventually. 

We expect our findings to encourage the study of magnetic phases for relevant quasi-1d materials. A promising platform to test our numerical results is Osmium Chloride (OsCl$_4$) a quasi-one dimensional material, in which the Os$^{4+}$ ions are in the $5d^4$ configuration. The model can further be adapted to describe systems in higher dimensions, e.g. $5d^4$ iridate, ruthenate and rhenate Mott insulators with a double perovskite structure.  

\section{acknowledgements}
We thank E. Miles Stoudenmire for help with the Intelligent Tensor Library (ITensor) open source code~\cite{ITensor}.
N.D.P. and N.T. acknowledge support from DOE grant DE-FG02-07ER46423. All computations were performed using the Unity cluster at the Ohio State University. J. H. H. acknowledge support from Samsung Science and Technology Foundation under Project Number SSTFBA1701-07.


\bibliography{reference.bib}

\begin{thebibliography}{10}
\expandafter\ifx\csname url\endcsname\relax
  \def\url#1{\texttt{#1}}\fi
\expandafter\ifx\csname urlprefix\endcsname\relax\def\urlprefix{URL }\fi
\providecommand{\bibinfo}[2]{#2}
\providecommand{\eprint}[2][]{\url{#2}}

\bibitem{trivedi15}
\bibinfo{author}{Meetei, O.~N.}, \bibinfo{author}{Cole, W.~S.},
  \bibinfo{author}{Randeria, M.} \& \bibinfo{author}{Trivedi, N.}
\newblock \bibinfo{title}{Novel magnetic state in ${d}^{4}$ mott insulators}.
\newblock \emph{\bibinfo{journal}{Phys. Rev. B}} \textbf{\bibinfo{volume}{91}},
  \bibinfo{pages}{054412} (\bibinfo{year}{2015}).
\newblock \urlprefix\url{http://link.aps.org/doi/10.1103/PhysRevB.91.054412}.

\bibitem{spinhall04}
\bibinfo{author}{Kato, Y.~K.}, \bibinfo{author}{Myers, R.~C.},
  \bibinfo{author}{Gossard, A.~C.} \& \bibinfo{author}{Awschalom, D.~D.}
\newblock \bibinfo{title}{Observation of the spin hall effect in
  semiconductors}.
\newblock \emph{\bibinfo{journal}{Science}} \textbf{\bibinfo{volume}{306}},
  \bibinfo{pages}{1910--1913} (\bibinfo{year}{2004}).
\newblock \urlprefix\url{http://science.sciencemag.org/content/306/5703/1910}.
\newblock
  \eprint{http://science.sciencemag.org/content/306/5703/1910.full.pdf}.

\bibitem{spinhall05}
\bibinfo{author}{Wunderlich, J.}, \bibinfo{author}{Kaestner, B.},
  \bibinfo{author}{Sinova, J.} \& \bibinfo{author}{Jungwirth, T.}
\newblock \bibinfo{title}{Experimental observation of the spin-hall effect in a
  two-dimensional spin-orbit coupled semiconductor system}.
\newblock \emph{\bibinfo{journal}{Phys. Rev. Lett.}}
  \textbf{\bibinfo{volume}{94}}, \bibinfo{pages}{047204}
  (\bibinfo{year}{2005}).
\newblock
  \urlprefix\url{http://link.aps.org/doi/10.1103/PhysRevLett.94.047204}.

\bibitem{qsh07}
\bibinfo{author}{K{\"o}nig, M.} \emph{et~al.}
\newblock \bibinfo{title}{Quantum spin hall insulator state in hgte quantum
  wells}.
\newblock \emph{\bibinfo{journal}{Science}} \textbf{\bibinfo{volume}{318}},
  \bibinfo{pages}{766--770} (\bibinfo{year}{2007}).
\newblock \urlprefix\url{http://science.sciencemag.org/content/318/5851/766}.
\newblock \eprint{http://science.sciencemag.org/content/318/5851/766.full.pdf}.

\bibitem{3d_ti}
\bibinfo{author}{Zhang, H.} \emph{et~al.}
\newblock \bibinfo{title}{Topological insulators in bi2se3, bi2te3 and sb2te3
  with a single dirac cone on the surface}.
\newblock \emph{\bibinfo{journal}{Nat Phys}} \textbf{\bibinfo{volume}{5}},
  \bibinfo{pages}{438--442} (\bibinfo{year}{2009}).
\newblock \urlprefix\url{http://dx.doi.org/10.1038/nphys1270}.

\bibitem{Khaliullin1}
\bibinfo{author}{Jackeli, G.} \& \bibinfo{author}{Khaliullin, G.}
\newblock \bibinfo{title}{{Mott} insulators in the strong spin-orbit coupling
  limit: From {Heisenberg} to a quantum compass and {Kitaev} models}.
\newblock \emph{\bibinfo{journal}{Phys. Rev. Lett.}}
  \textbf{\bibinfo{volume}{102}}, \bibinfo{pages}{017205}
  (\bibinfo{year}{2009}).
\newblock
  \urlprefix\url{https://link.aps.org/doi/10.1103/PhysRevLett.102.017205}.

\bibitem{kim08}
\bibinfo{author}{Kim, B.~J.} \emph{et~al.}
\newblock \bibinfo{title}{Novel ${J}_{\mathrm{eff}}=1/2$ mott state induced by
  relativistic spin-orbit coupling in ${\mathrm{sr}}_{2}{\mathrm{iro}}_{4}$}.
\newblock \emph{\bibinfo{journal}{Phys. Rev. Lett.}}
  \textbf{\bibinfo{volume}{101}}, \bibinfo{pages}{076402}
  (\bibinfo{year}{2008}).
\newblock
  \urlprefix\url{http://link.aps.org/doi/10.1103/PhysRevLett.101.076402}.

\bibitem{kim09}
\bibinfo{author}{Kim, B.~J.} \emph{et~al.}
\newblock \bibinfo{title}{Phase-sensitive observation of a spin-orbital mott
  state in sr2iro4}.
\newblock \emph{\bibinfo{journal}{Science}} \textbf{\bibinfo{volume}{323}},
  \bibinfo{pages}{1329--1332} (\bibinfo{year}{2009}).
\newblock \urlprefix\url{http://science.sciencemag.org/content/323/5919/1329}.
\newblock
  \eprint{http://science.sciencemag.org/content/323/5919/1329.full.pdf}.

\bibitem{balents10}
\bibinfo{author}{Pesin, D.} \& \bibinfo{author}{Balents, L.}
\newblock \bibinfo{title}{Mott physics and band topology in materials with
  strong spin{\textendash}orbit interaction}.
\newblock \emph{\bibinfo{journal}{Nat Phys}} \textbf{\bibinfo{volume}{6}},
  \bibinfo{pages}{376--381} (\bibinfo{year}{2010}).
\newblock \urlprefix\url{http://dx.doi.org/10.1038/nphys1606}.

\bibitem{balents10-d1}
\bibinfo{author}{Chen, G.}, \bibinfo{author}{Pereira, R.} \&
  \bibinfo{author}{Balents, L.}
\newblock \bibinfo{title}{Exotic phases induced by strong spin-orbit coupling
  in ordered double perovskites}.
\newblock \emph{\bibinfo{journal}{Phys. Rev. B}} \textbf{\bibinfo{volume}{82}},
  \bibinfo{pages}{174440} (\bibinfo{year}{2010}).
\newblock \urlprefix\url{http://link.aps.org/doi/10.1103/PhysRevB.82.174440}.

\bibitem{balents11-d2}
\bibinfo{author}{Chen, G.} \& \bibinfo{author}{Balents, L.}
\newblock \bibinfo{title}{Spin-orbit coupling in ${d}^{2}$ ordered double
  perovskites}.
\newblock \emph{\bibinfo{journal}{Phys. Rev. B}} \textbf{\bibinfo{volume}{84}},
  \bibinfo{pages}{094420} (\bibinfo{year}{2011}).
\newblock \urlprefix\url{http://link.aps.org/doi/10.1103/PhysRevB.84.094420}.

\bibitem{trivedi13}
\bibinfo{author}{Meetei, O.~N.}, \bibinfo{author}{Erten, O.},
  \bibinfo{author}{Randeria, M.}, \bibinfo{author}{Trivedi, N.} \&
  \bibinfo{author}{Woodward, P.}
\newblock \bibinfo{title}{Theory of high ${T}_{c}$ ferrimagnetism in a
  multiorbital mott insulator}.
\newblock \emph{\bibinfo{journal}{Phys. Rev. Lett.}}
  \textbf{\bibinfo{volume}{110}}, \bibinfo{pages}{087203}
  (\bibinfo{year}{2013}).
\newblock
  \urlprefix\url{http://link.aps.org/doi/10.1103/PhysRevLett.110.087203}.

\bibitem{Trivedi1_d4}
\bibinfo{author}{Svoboda, C.}, \bibinfo{author}{Randeria, M.} \&
  \bibinfo{author}{Trivedi, N.}
\newblock \bibinfo{title}{Effective magnetic interactions in spin-orbit coupled
  ${d}^{4}$ mott insulators}.
\newblock \emph{\bibinfo{journal}{Phys. Rev. B}} \textbf{\bibinfo{volume}{95}},
  \bibinfo{pages}{014409} (\bibinfo{year}{2017}).
\newblock \urlprefix\url{https://link.aps.org/doi/10.1103/PhysRevB.95.014409}.

\bibitem{Nitin1}
\bibinfo{author}{Kaushal, N.} \emph{et~al.}
\newblock \bibinfo{title}{Density matrix renormalization group study of a
  three-orbital hubbard model with spin-orbit coupling in one dimension}.
\newblock \emph{\bibinfo{journal}{Phys. Rev. B}} \textbf{\bibinfo{volume}{96}},
  \bibinfo{pages}{155111} (\bibinfo{year}{2017}).
\newblock \urlprefix\url{https://link.aps.org/doi/10.1103/PhysRevB.96.155111}.

\bibitem{Nitin2}
\bibinfo{author}{Kaushal, N.}, \bibinfo{author}{Nocera, A.},
  \bibinfo{author}{Alvarez, G.}, \bibinfo{author}{Moreo, A.} \&
  \bibinfo{author}{Dagotto, E.}
\newblock \bibinfo{title}{Block excitonic condensate at $n=3.5$ in a spin-orbit
  coupled ${t}_{2g}$ multiorbital hubbard model}.
\newblock \emph{\bibinfo{journal}{Phys. Rev. B}} \textbf{\bibinfo{volume}{99}},
  \bibinfo{pages}{155115} (\bibinfo{year}{2019}).
\newblock \urlprefix\url{https://link.aps.org/doi/10.1103/PhysRevB.99.155115}.

\bibitem{CaRuO_1}
\bibinfo{author}{Nakatsuji, S.}, \bibinfo{author}{Ikeda, S.-i.} \&
  \bibinfo{author}{Maeno, Y.}
\newblock \bibinfo{title}{{Ca$_2$RuO$_4$}: New {Mott} insulators of layered
  ruthenate}.
\newblock \emph{\bibinfo{journal}{Journal of the Physical Society of Japan}}
  \textbf{\bibinfo{volume}{66}}, \bibinfo{pages}{1868--1871}
  (\bibinfo{year}{1997}).
\newblock \urlprefix\url{https://doi.org/10.1143/JPSJ.66.1868}.
\newblock \eprint{https://doi.org/10.1143/JPSJ.66.1868}.

\bibitem{CaRuO_2}
\bibinfo{author}{Braden, M.}, \bibinfo{author}{Andr\'e, G.},
  \bibinfo{author}{Nakatsuji, S.} \& \bibinfo{author}{Maeno, Y.}
\newblock \bibinfo{title}{Crystal and magnetic structure of
  ${\mathrm{ca}}_{2}{\mathrm{ruo}}_{4}:$ magnetoelastic coupling and the
  metal-insulator transition}.
\newblock \emph{\bibinfo{journal}{Phys. Rev. B}} \textbf{\bibinfo{volume}{58}},
  \bibinfo{pages}{847--861} (\bibinfo{year}{1998}).
\newblock \urlprefix\url{https://link.aps.org/doi/10.1103/PhysRevB.58.847}.

\bibitem{CaRuO_3}
\bibinfo{author}{Carlo, J.} \emph{et~al.}
\newblock \bibinfo{title}{New magnetic phase diagram of {(Sr,
  Ca)}$_{2}$ruo$_{4}$}.
\newblock \emph{\bibinfo{journal}{Nature materials}}
  \textbf{\bibinfo{volume}{11}}, \bibinfo{pages}{323} (\bibinfo{year}{2012}).
\newblock \urlprefix\url{https://doi.org/10.1038/nmat3236}.

\bibitem{PeroSkite_1}
\bibinfo{author}{Cao, G.} \emph{et~al.}
\newblock \bibinfo{title}{Novel magnetism of ${\mathrm{ir}}^{5+}(5{d}^{4})$
  ions in the double perovskite ${\mathrm{sr}}_{2}{\mathrm{yiro}}_{6}$}.
\newblock \emph{\bibinfo{journal}{Phys. Rev. Lett.}}
  \textbf{\bibinfo{volume}{112}}, \bibinfo{pages}{056402}
  (\bibinfo{year}{2014}).
\newblock
  \urlprefix\url{https://link.aps.org/doi/10.1103/PhysRevLett.112.056402}.

\bibitem{PeroSkite_2}
\bibinfo{author}{Laguna-Marco, M.~A.} \emph{et~al.}
\newblock \bibinfo{title}{Electronic structure, local magnetism, and spin-orbit
  effects of ir(iv)-, ir(v)-, and ir(vi)-based compounds}.
\newblock \emph{\bibinfo{journal}{Phys. Rev. B}} \textbf{\bibinfo{volume}{91}},
  \bibinfo{pages}{214433} (\bibinfo{year}{2015}).
\newblock \urlprefix\url{https://link.aps.org/doi/10.1103/PhysRevB.91.214433}.

\bibitem{PeroSkite_3}
\bibinfo{author}{Bhowal, S.}, \bibinfo{author}{Baidya, S.},
  \bibinfo{author}{Dasgupta, I.} \& \bibinfo{author}{Saha-Dasgupta, T.}
\newblock \bibinfo{title}{Breakdown of $j=0$ nonmagnetic state in ${d}^{4}$
  iridate double perovskites: A first-principles study}.
\newblock \emph{\bibinfo{journal}{Phys. Rev. B}} \textbf{\bibinfo{volume}{92}},
  \bibinfo{pages}{121113} (\bibinfo{year}{2015}).
\newblock \urlprefix\url{https://link.aps.org/doi/10.1103/PhysRevB.92.121113}.

\bibitem{BaYIrO_1}
\bibinfo{author}{Fuchs, S.} \emph{et~al.}
\newblock \bibinfo{title}{Unraveling the nature of magnetism of the $5{d}^{4}$
  double perovskite ${\mathrm{ba}}_{2}{\mathrm{yiro}}_{6}$}.
\newblock \emph{\bibinfo{journal}{Phys. Rev. Lett.}}
  \textbf{\bibinfo{volume}{120}}, \bibinfo{pages}{237204}
  (\bibinfo{year}{2018}).
\newblock
  \urlprefix\url{https://link.aps.org/doi/10.1103/PhysRevLett.120.237204}.

\bibitem{BaYIrO_2}
\bibinfo{author}{Ranjbar, B.} \emph{et~al.}
\newblock \bibinfo{title}{Structural and magnetic properties of the iridium
  double perovskites ba2–xsrxyiro6}.
\newblock \emph{\bibinfo{journal}{Inorganic Chemistry}}
  \textbf{\bibinfo{volume}{54}}, \bibinfo{pages}{10468--10476}
  (\bibinfo{year}{2015}).
\newblock \urlprefix\url{https://doi.org/10.1021/acs.inorgchem.5b01905}.

\bibitem{BaYIrO_3}
\bibinfo{author}{Phelan, B.~F.}, \bibinfo{author}{Seibel, E.~M.},
  \bibinfo{author}{Badoe~Jr, D.}, \bibinfo{author}{Xie, W.} \&
  \bibinfo{author}{Cava, R.~J.}
\newblock \bibinfo{title}{Influence of structural distortions on the ir
  magnetism in ba2- xsrxyiro6 double perovskites}.
\newblock \emph{\bibinfo{journal}{Solid State Communications}}
  \textbf{\bibinfo{volume}{236}}, \bibinfo{pages}{37--40}
  (\bibinfo{year}{2016}).

\bibitem{BaYIrO_4}
\bibinfo{author}{Dey, T.} \emph{et~al.}
\newblock \bibinfo{title}{${\text{ba}}_{2}{\text{yiro}}_{6}$: A cubic double
  perovskite material with ${\text{ir}}^{5+}$ ions}.
\newblock \emph{\bibinfo{journal}{Phys. Rev. B}} \textbf{\bibinfo{volume}{93}},
  \bibinfo{pages}{014434} (\bibinfo{year}{2016}).
\newblock \urlprefix\url{https://link.aps.org/doi/10.1103/PhysRevB.93.014434}.

\bibitem{Ruthnate_1}
\bibinfo{author}{Kumar, R.} \emph{et~al.}
\newblock \bibinfo{title}{Unconventional magnetism in the $4{d}^{4}$-based
  $s=1$ honeycomb system
  ${\mathrm{ag}}_{3}{\mathrm{liru}}_{2}{\mathrm{o}}_{6}$}.
\newblock \emph{\bibinfo{journal}{Phys. Rev. B}} \textbf{\bibinfo{volume}{99}},
  \bibinfo{pages}{054417} (\bibinfo{year}{2019}).
\newblock \urlprefix\url{https://link.aps.org/doi/10.1103/PhysRevB.99.054417}.

\bibitem{Ruthnate_2}
\bibinfo{author}{Wang, J.~C.} \emph{et~al.}
\newblock \bibinfo{title}{Lattice-tuned magnetism of
  ${\mathrm{ru}}^{4+}(4{d}^{4})$ ions in single crystals of the layered
  honeycomb ruthenates ${\mathrm{li}}_{2}{\mathrm{ruo}}_{3}$ and
  ${\mathrm{na}}_{2}{\mathrm{ruo}}_{3}$}.
\newblock \emph{\bibinfo{journal}{Phys. Rev. B}} \textbf{\bibinfo{volume}{90}},
  \bibinfo{pages}{161110} (\bibinfo{year}{2014}).
\newblock \urlprefix\url{https://link.aps.org/doi/10.1103/PhysRevB.90.161110}.

\bibitem{Ruthnate_3}
\bibinfo{author}{Gapontsev, V.~V.} \emph{et~al.}
\newblock \bibinfo{title}{Spectral and magnetic properties of na2ruo3}.
\newblock \emph{\bibinfo{journal}{Journal of Physics: Condensed Matter}}
  \textbf{\bibinfo{volume}{29}}, \bibinfo{pages}{405804}
  (\bibinfo{year}{2017}).
\newblock \urlprefix\url{https://doi.org/10.1088/1361-648x/aa7fd6}.

\bibitem{ULS1}
\bibinfo{author}{Sutherland, B.}
\newblock \bibinfo{title}{Model for a multicomponent quantum system}.
\newblock \emph{\bibinfo{journal}{Phys. Rev. B}} \textbf{\bibinfo{volume}{12}},
  \bibinfo{pages}{3795} (\bibinfo{year}{1975}).
\newblock \urlprefix\url{https://doi.org/10.1103/PhysRevB.12.3795}.

\bibitem{ULS2}
\bibinfo{author}{Uimin, G.~V.}
\newblock \bibinfo{title}{One-dimensional problem for s = 1 with modified
  antiferromagnetic hamiltonian}.
\newblock \emph{\bibinfo{journal}{ZhETF Pis. Red.}}
  \textbf{\bibinfo{volume}{12}}, \bibinfo{pages}{225} (\bibinfo{year}{1970}).
\newblock
  \urlprefix\url{http://www.jetpletters.ac.ru/ps/1730/article_26296.shtml}.

\bibitem{ULS3}
\bibinfo{author}{Lai, C.~K.}
\newblock \bibinfo{title}{Lattice gas with nearest‐neighbor interaction in
  one dimension with arbitrary statistics}.
\newblock \emph{\bibinfo{journal}{J. Math. Phys.}}
  \textbf{\bibinfo{volume}{15}}, \bibinfo{pages}{1675} (\bibinfo{year}{1974}).
\newblock \urlprefix\url{https://doi.org/10.1063/1.1666522}.

\bibitem{ITensor}
\emph{\bibinfo{journal}{\mbox{ITensor Library} (version 2.0.11)
  http://itensor.org}} .

\bibitem{littlewood98}
\bibinfo{author}{F\'ath, G.} \& \bibinfo{author}{Littlewood, P.~B.}
\newblock \bibinfo{title}{Massless phases of haldane-gap antiferromagnets in a
  magnetic field}.
\newblock \emph{\bibinfo{journal}{Phys. Rev. B}} \textbf{\bibinfo{volume}{58}},
  \bibinfo{pages}{R14709--R14712} (\bibinfo{year}{1998}).
\newblock \urlprefix\url{http://link.aps.org/doi/10.1103/PhysRevB.58.R14709}.

\end{thebibliography}

\begin{widetext}
\clearpage
\begin{center}
\textbf{\large Supplemental: Magnetic phase transitions in quantum spin-orbital liquids}
\end{center}
\end{widetext}

\setcounter{equation}{0}
\setcounter{figure}{0}
\setcounter{table}{0}
\setcounter{page}{1}
\setcounter{section}{0}
\makeatletter
\renewcommand{\theequation}{S\arabic{equation}}
\renewcommand{\thefigure}{S\arabic{figure}}
\renewcommand{\bibnumfmt}[1]{[S#1]}

\section{Operators and Observables}
In this section, we define all the observables that are used in the main text. Figure 2(b,d) shows the total quantum number in different sectors, i.e. the measure of the total $S, \; L, \; J$ sector in the g.s., which are defined as $S_T, \; L_T, \; J_T$:
\begin{equation} \label{eq:S1}
    \begin{split}
        S_T^2 &= S_T (S_T + 1) = \sum_{i,j} \< \mathbf{S}_i \cdot \mathbf{S}_j \> \\
        L_T^2 &= L_T (L_T + 1) = \sum_{i,j} \< \mathbf{L}_i \cdot \mathbf{L}_j \> \\
        J_T^2 &= J_T (J_T + 1) = \sum_{i,j} \< \mathbf{J}_i \cdot \mathbf{J}_j \> \\
            &= S_T + L_T + 2 \sum_{i,j} \< \mathbf{S}_i \cdot \mathbf{L}_j \>
    \end{split}
\end{equation}
where $\<\cdot\>$ denotes the expectation of the g.s. wavefunction, and index $i$ runs over all sites in the chain. Figure 2(c) shows the $\hat{z}$ component of total spin, orbital and J magnetization. They are defined as: 

\begin{equation}
    \begin{split}
        \< \bar{S^z} \> &= \sum_{i} \< S^z_i \> \\
        \< \bar{L^z} \> &= \sum_{i} \< L^z_i \> \\
        \< \bar{J^z} \> &= \< \bar{S^z} \> + \< \bar{L^z} \>
    \end{split}
\end{equation}
In the main text, we also performed calculation on (connected) correlation function on different quantum sectors, both in real space and momentum space (Figure 3). the (connected) correlation of two observables $\textbf{A}$ and $\textbf{B}$ on site $i,j$, which can be interpreted as a measure of correlation in fluctuation, is:

\begin{equation}
    \< \mathbf{\delta A}_{i} \cdot \mathbf{\delta B}_{j} \> = \< \mathbf{A}_i \cdot \mathbf{B}_j \> - \< \mathbf{A}_i \> \cdot \< \mathbf{B}_j \>
\end{equation}
Let $N_R$ be the total number of pairs at distance $R$, where $R=|i-j|$. The real space correlation at fixed distance $R$ is then defined as:

\begin{equation}
    \begin{split}
        C_S(R) &= \frac{1}{N_R} \sum_{R} \< \mathbf{\delta S}_{i} \cdot \mathbf{\delta S}_{i+R} \> \\
        C_L(R) &= \frac{1}{N_R} \sum_{R} \< \mathbf{\delta L}_{i} \cdot \mathbf{\delta L}_{i+R} \> \\
        C_J(R) &= \frac{1}{N_R} \sum_{R} \< \mathbf{\delta J}_{i} \cdot \mathbf{\delta J}_{i+R} \> 
    \end{split}
\end{equation}
where the $J-J$ correlation can be written explicitly as:
\begin{equation}
    \begin{split}
        \< \mathbf{\delta J}_{i} \cdot \mathbf{\delta J}_{j} \> &= \< \mathbf{\delta S}_i \cdot \mathbf{\delta S}_j \> + \< \mathbf{\delta L}_i \cdot \mathbf{\delta L}_j \> \\
        &+ 2 \ \< \mathbf{\delta S}_i \cdot \mathbf{\delta L}_j \> 
    \end{split}
\end{equation}

In their momentum space, we have:

\begin{equation}
    \begin{split}
        S(k) &= \frac{1}{N^2} \sum_{i,j} e^{ik(r_i - r_j)} \< \mathbf{\delta S}_{i} \cdot \mathbf{\delta S}_{j} \>  \\
        L(k) &= \frac{1}{N^2} \sum_{i,j} e^{ik(r_i - r_j)} \< \mathbf{\delta L}_{i} \cdot \mathbf{\delta L}_{j} \>  \\
        S(k) &= \frac{1}{N^2} \sum_{i,j} e^{ik(r_i - r_j)} \< \mathbf{\delta J}_{i} \cdot \mathbf{\delta J}_{j} \>  \\
    \end{split}
\end{equation}
In Figure 3(g,h,i), in order to better visualize the correlation in real space, we defined $G_S(R),\;G_L(R),\;G_J(R)$ as the real-space alignment in different quantum sectors with respect to a site of reference $i_r$:
\begin{equation}
    \begin{split}
        G_S(R) &= \< \mathbf{\delta S}_{i_r} \cdot \mathbf{\delta S}_{i_r+R} \> \\
        G_L(R) &= \< \mathbf{\delta L}_{i_r} \cdot \mathbf{\delta L}_{i_r+R} \> \\
        G_J(R) &= \< \mathbf{\delta J}_{i_r} \cdot \mathbf{\delta J}_{i_r+R} \> 
    \end{split}
\end{equation}

\section{Building blocks of $H_{d4}$ Hamiltonian}
\subsection{Projection Operator }

To articulate the structure of the $H_{d4}$ Hamiltonian and understand the correlation profile at $\lambda = 0$ in the maintext, we study several building blocks of the $H_{d4}$ model in terms of the projection operator i.e. the Heisenberg, AKLT and ULS Hamiltonian.

A projector $\textit{P}^{(n)}$ is in general an operator that squares to itself, with eigenvalues 0 or 1. Therefore, any Hamiltonian in the form of $H = \sum_i \textit{P}^{(i)}$ has non-negative eigenvalues
\begin{equation}
\begin{split}
    &\textit{P}^{(n)} \textit{P}^{(n)} = \textit{P}^{(n)}\\
    &E_{n} \geq 0 
\end{split}
\end{equation}

Let $\textit{P}_{\<ij\>}^{(m)}$ be the projection operator acting on a local spin dimer with neighboring end points $(i,j)$, whose eigenvalue is $E_i = \delta_{i,m}$:
\begin{equation} \label{eq:S8}
    \begin{split}
        \textit{P}^{(n)}_{\<ij\>}|S_T = m\> = \delta_{n,m}|S_T = m\>\\
    \end{split}
\end{equation}
and all these projection operators add up to identity:
\begin{equation} \label{eq:identity}
    \sum_{n}\textit{P}^{(n)}_{\<ij\>} = 1
\end{equation}
This makes the $S_z = m$ state energetically unfavorable compared to others, thus, effectively tend to annihilate $S_z = m$ in low energy regime.

\subsection{Heisenberg Hamiltonian in the form of projection operators}
Consider a standard spin-1/2 Heisenberg chain with only exchange interaction. First let's build up projection operators in the local dimer Hilbert space:  $\textit{P}^{(1)}_{\<ij\>}$ and $\textit{P}^{(0)}_{\<ij\>}$, which respectively denotes projection to $S_T = 1,0$. The quantum number $S_T$ is given by (Eq.~\ref{eq:S1}), in which  $\textbf{S}_T^2 = \sum_{i,j}\textbf{S}_i\cdot\textbf{S}_j$ before evaluation. This operator gives $S_T(S_T+1)=2$ when acting on triplet state, and $S_T(S_T+1)=0$ on singlet state. Therefore we construct the projection operator as such so that they satisfies (Eq.~\ref{eq:S8}): 
\begin{equation}
    \begin{split}
        &\textit{P}^{(1)}_{\<ij\>}=\frac{1}{2}\textbf{S}_T^2\\
        &\textit{P}^{(0)}_{\<ij\>}=-\frac{1}{2}(\textbf{S}_T^2-2)        
    \end{split}
\end{equation}
Note that the definition above also satisfies the identity relation: $\sum_n P^{(n)}_{\<ij\>} = 1$.
Now, add up projection operators on each pair of sites on the chain with exchange constant $J$. First we add up all $\textit{P}^{(0)}$.
\begin{equation}
    \begin{split}
        H_{FM} &=J\sum_{\<ij\>}\textit{P}^{(0)}_{\<ij\>}\\
        &= -J\sum_{\<ij\>}\textbf{S}_i\cdot\textbf{S}_j -\frac{1}{4}
    \end{split}
\end{equation}
With $J > 0$, this is just the ferromagnetic Heisenberg Hamiltonian with a $\frac{1}{4}$ energy shift.

By the same logic, we get the anti-ferromagnetic Hamiltonian by adding up $\textit{P}^{(1)}_{\<ij\>}$ that favors local singlet states across all sites:

\begin{equation}
    \begin{split}
        H_{AFM} &=J\sum_{\<ij\>}\textit{P}^{(1)}_{\<ij\>}\\
        &= J\sum_{\<ij\>}\textbf{S}_i\cdot\textbf{S}_j + \frac{3}{4}
    \end{split}
\end{equation}

Hamiltonians written in projection operators provide intuition of local faverable states, it is also straightforward to look at the frustration by numerical simulation. The ground states calculated from the Hamiltonian will produce $E_{g.s.} = 0$ if the system is frustration-free that the energy of all local bonds can be simultaneously projected out (minimized), and a $E_{g.s.} > 0$ is indicative of the presence of frustration.

\subsection{AKLT Hamiltonian}
The AKLT Hamiltonian is essentially a model of spin-1/2 chain with singlet bond connecting pairs of sites. The irreducible representation of 2 neighboring pairs is:
\begin{equation}
    \frac{1}{2}\otimes 0 \otimes \frac{1}{2} = 1 \oplus 0
\end{equation}

The same Hilbert space can be spanned in a spin-1 chain, with two spin-1/2 sites make up a spin-1 site, as showed in figure 1.
\begin{figure}
        \center
	\includegraphics[width=0.5\textwidth]{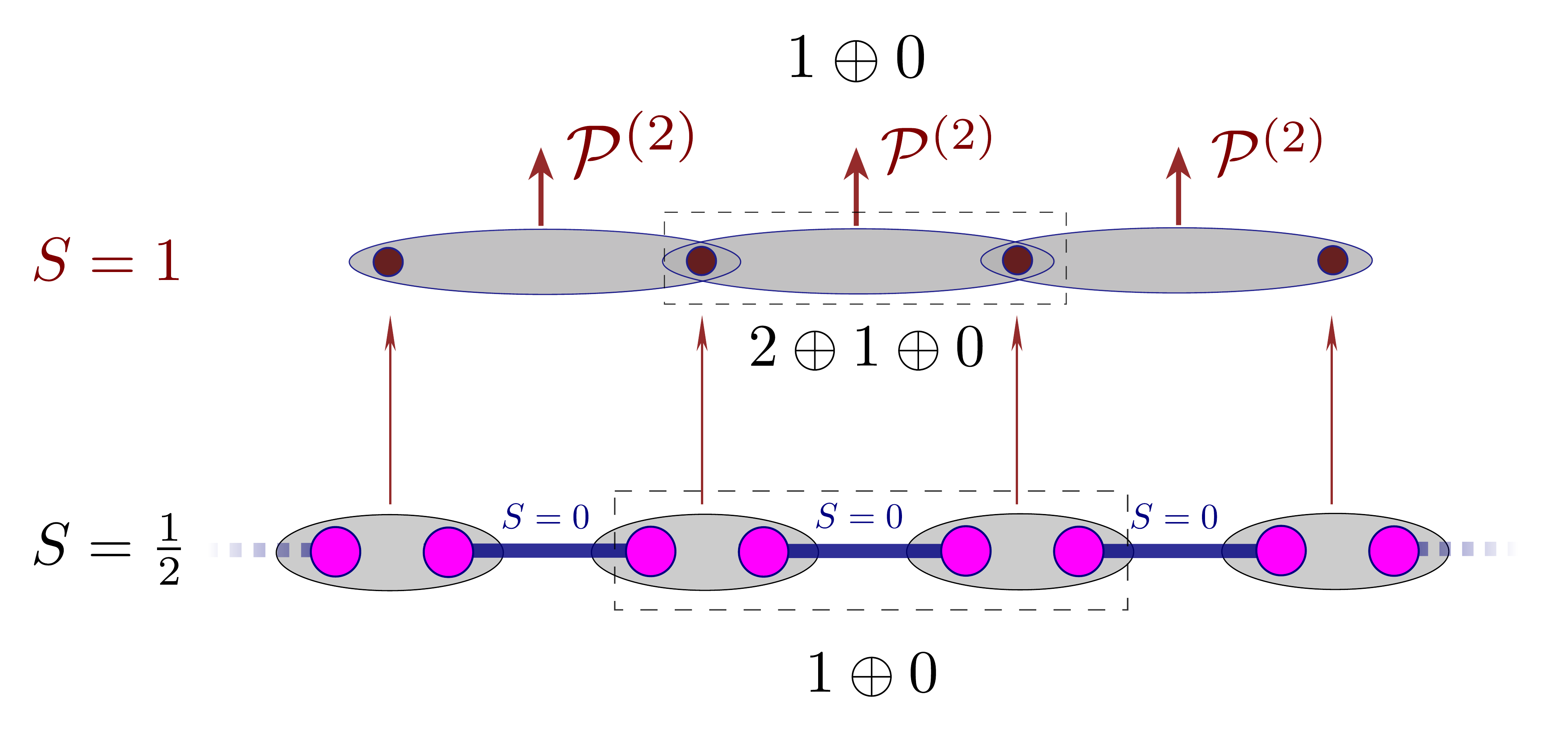}
        \caption{The AKLT model is built from the state where spin- 1/2 pairs form singlets (S = 0) between adjacent sites. This is essentially a spin-1 chain with a non-trivial frustration that minimize s=2 configuration}
\end{figure}
The Hilbert space of 2 neighboring spin-1 sites is presented by:
\begin{equation}
    1 \otimes 1 = 2 \oplus 1 \oplus 0
\end{equation}

The first "2" on R.H.S. has to be ruled out to fully recover the subspace of the original spin-1/2 model. Therefore, based on the argument in last subsection, we add up projection operators which have eigenvalue $ E_{S_T} = \delta_{S_T,2} $. That is :
\begin{equation}
	 H_{AKLT} = \sum_{\<ij\>} \textit{P}^{(2)}_{\<ij\>}
\end{equation}

The ground state of this Hamiltonian locally favors $S_T=0,1$ and tend to rule out $S_T = 2$ bonds (can be projected out completely if frustration-free), leaving only $1\oplus 0$ subspace.
3 projection operators are:
\begin{equation}
    \begin{split}
        	\textit{P}^{(2)}_{\<ij\>} &= \frac{1}{24}\textbf{S}_T^2(\textbf{S}_T^2-2)\\
        	\textit{P}^{(1)}_{\<ij\>} &= -\frac{1}{8}\textbf{S}_T^2(\textbf{S}_T^2-6)\\
        	\textit{P}^{(0)}_{\<ij\>} &= \frac{1}{12}(\textbf{S}_T^2-2)(\textbf{S}_T^2-6)
    \end{split}
\end{equation}

By writing out these operator explicitly in spin-1 case, similarly, by $\textbf{S}_T^2 = \sum_{i,j}\textbf{S}_i\cdot\textbf{S}_j$ where $i,j$ run over 2 neighboring sites, the AKLT Hamiltonian can be expanded as:
\begin{equation}
    \begin{split}
        H_{AKLT} &= \sum_{\<ij\>}\textit{P}^{(2)}_{\<ij\>}\\
        &=\sum_{\<ij\>}\frac{1}{24}\textbf{S}_T^2(\textbf{S}_T^2-2)\\
        &=\sum_{\<ij\>}(\textbf{S}_i\textbf{S}_j)+\frac{1}{3}(\textbf{S}_i\textbf{S}_j)^2+\frac{2}{3}
    \end{split}
\end{equation}
This is the standard AKLT Hamiltonian.
\begin{figure}
        \center
	\includegraphics[width=0.5\textwidth]{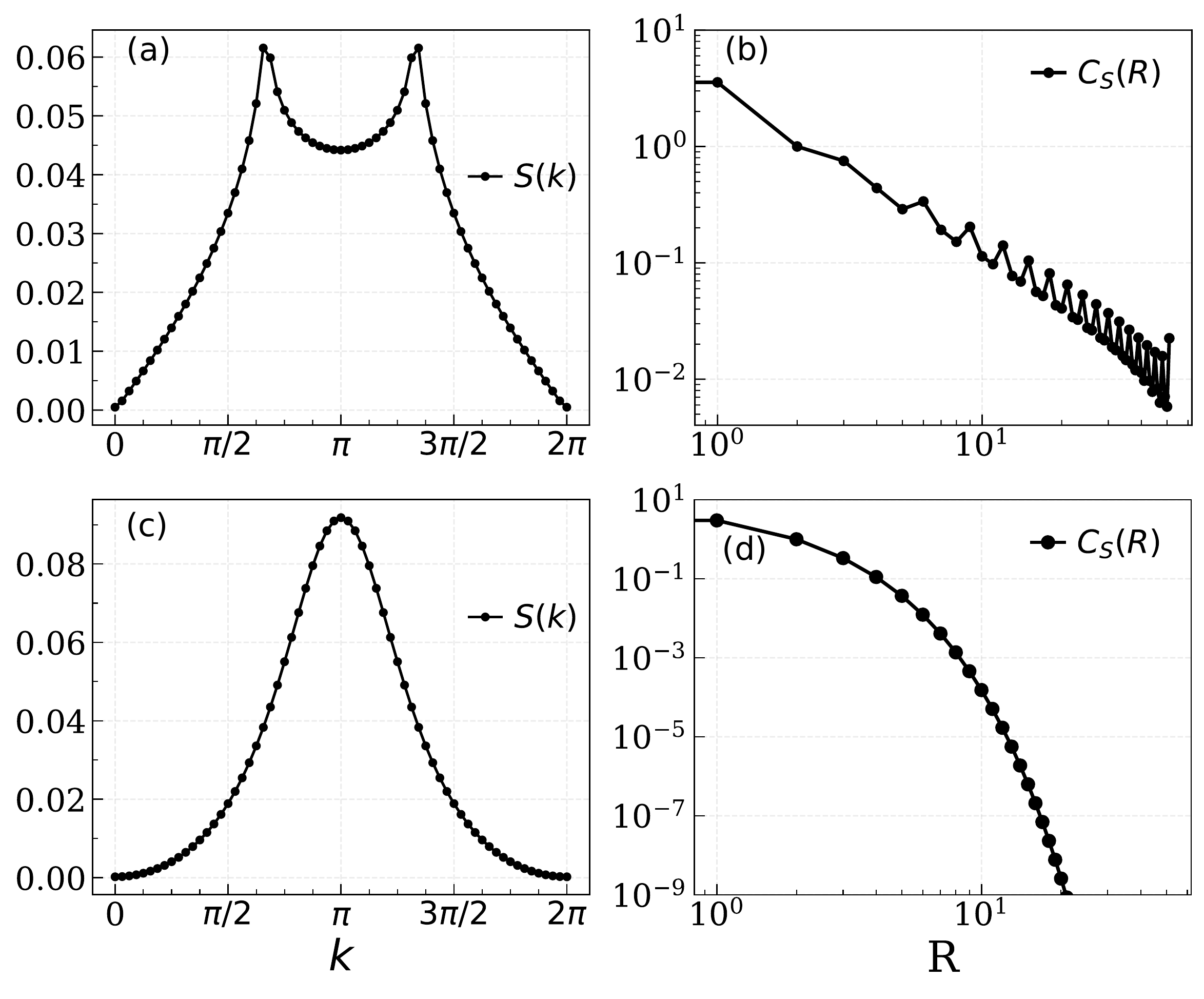}
        \caption{ (a,c) momentum-space correlations,  and (b,d) decay of real-space correlations of ULS and AKLT respectively. Results are obtained using 64 sites DMRG simulation}
\end{figure}
\subsection{ULS Hamiltonian}
As is argued in the main text, the local ULS Hamiltonian favors $S_T = 1$ bonds. This property can be achieved by projection operator $-\textit{P}^{(1)}_{\<ij\>}$ where the minus sign makes $S_T = 1$ bonds energetically favorable. The full ULS Hamiltonian is then described by $H_{ULS} = - \sum_{\<ij\>} \textit{P}^{(1)}_{\<ij\>}$.
Write out the projection operator explicitly:
\begin{equation}
    \begin{split}
        H_{ULS} &= \sum_{\<ij\>}-\textit{P}^{(1)}_{\<ij\>}\\
        &= \sum_{\<ij\>}\frac{1}{8}\textbf{S}_T^2(\textbf{S}_T^2-6)\\
        &=\frac{1}{2} \sum_{\<ij\>} (\textbf{S}_i\textbf{S}_j)^2 +(\textbf{S}_i \textbf{S}_j) -2
    \end{split}
\end{equation}

By writing the identity relation (Eq.~\ref{eq:identity}) explicitly, the ULS Hamiltonian can be written in terms of two competing projection operators defined in spin-1 Hilbert space:
\begin{equation} \label{eq:ULS2}
    H_{ULS} = \sum_{\<ij\>} \textit{P}^{(2)}_{\<ij\>} +\textit{P}^{(0)}_{\<ij\>}-1
\end{equation}
where the first projection operator in the summation is the AKLT term. (Eq.~\ref{eq:ULS2}) provides a qualitative explanation for the bifurcation in the $S(k)$ in the g.s. of ULS Hamiltonian.

Qualitatively, the first $\textit{P}^{(2)}$ term, i.e. $h_{AKLT}$, in Hamiltonian prevents the neighboring sites from being aligned with each other, it prefers a valance bond solid state in the auxilliary spin-1/2 chain, thus prevents the formation of a global $\ket{FM}$ state. Alone with this term, the Hamiltonian is frustration-free, thus the AKLT chain can reach its minimal energy at ground state locally. DMRG calculation also shows that the ground state eigenvalue is zero. This rules out the possibility of forming a ferromagnetic order(shown in Figure.S2), and results in an AFM-like state with no contribution from $k = 0$ and $2\pi$. 

The frustration occurs in the ULS Hamiltonian. This can be done by adding the second term $\textit{P}^{(0)}_{\<ij\>}$ into the AKLT Hamiltonian, and we are back at (Eq.~\ref{eq:ULS2}). Again this is verified by DMRG calculation which shows a non-zero eigenvalue at the ground state. Since $\textit{P}^{(0)}_{\<ij\>}$ avoids local singlet state, the peak at $k=\pi$ will be lowered compared to an AKLT ground state. This quialitatively explains the double incommensurate peaks in momentum space correlations at $\lambda = 0$ and S1 phase in the main text.

\section{Symmetries}
The orbital sector of the $H_{d4}$ (Eq.~\ref{eq:hamiltonian}) is well-known as the ULS Hamiltonian, which also occurs in the ULSZ approximation in section IV. This is known to exhibit a SU(3) symmetry. In this section, beginning with the simplest spin-1/2 Heisenberg Hamiltonian, we discuss the fermionic representation of ULS Hamiltonian relavent for both $H_{d4}$ and $H_{ULSZ}$, which helps to understand if certain system has larger symmetry than it seems.
In spin-1/2 Heisenberg Hamiltonian, define the spin-1/2 operator $\textbf{S}_i \equiv \frac{1}{2} \psi_i^\dagger\vec{\sigma}\psi_i$, $\psi_i = (c_{i,\uparrow},c_{i,\downarrow})$, where $c_{i,\uparrow}(c_{i,\downarrow})$ annihilates (creates) a fermion with spin  $\sigma$ at site i. we have:
\begin{equation}
	\begin{split}
		S^z_i &= \frac{1}{2}(c_{i\uparrow}^\dagger c_{i\uparrow}-c_{i\downarrow}^\dagger c_{i\downarrow})\\
		S^+ &= c_{i\uparrow}^\dagger c_{i\downarrow},\; S^- = (S^+)^\dagger
	\end{split}
\end{equation}

There is a constraint that the total spin on every site is  $\frac{1}{2}$. Therefore:
\begin{equation}
	\begin{split}
		\textbf{S}_i^2 &= S_i^z S_i^z +\frac{1}{2}(S^+_i S^-_i + S^-_i S^+_i)\\
			       &=\frac{3}{4}(n_i -2 n_{i,\uparrow}n_{i\downarrow})\\
			       &= \frac{3}{4}(n_{i\uparrow}- n_{i\downarrow})^2 \equiv \frac{3}{4}
	\end{split}
\end{equation}
where $n_i = n_{i,\uparrow}+n_{i,\downarrow}$ is the fermion number operator at site i. In the last step we set  $n_{i,\uparrow}+n_{i,\downarrow} = \pm 1$, thus enforce the constraint that there's only one fermion on each site.

In the full Hilbert space spanned by two sites $(i,j)$, we have:
\begin{equation}
	\begin{split}
        I = n_i n_j
	\end{split}
\end{equation}

\begin{figure*}[t]
  \centering
    \includegraphics[trim={0cm 0cm -0.0cm 0cm},width=0.98\textwidth]{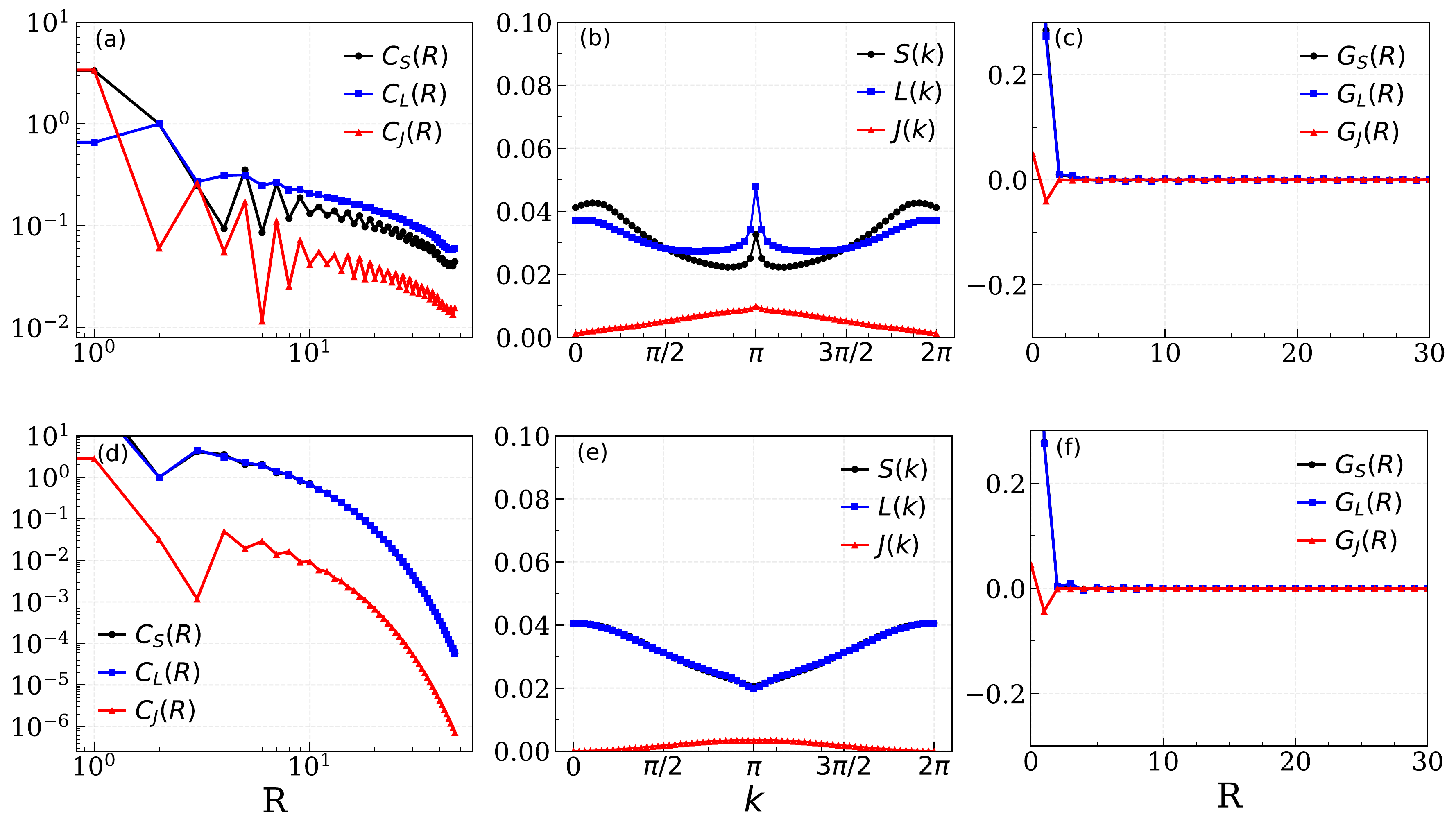}
    \subfloat{\label{fig:3a}} \subfloat{\label{fig:3b}} \subfloat{\label{fig:3c}} \subfloat{\label{fig:3d}} \subfloat{\label{fig:3e}} \subfloat{\label{fig:3f}} 
    \caption{
    (a,c) Decay of the spin, orbital and total angular momentum $J$ correlations (Eq.~\ref{eq:RealSpCorr}) near the S1-SOS phase transition, $\lambda = 1.19 $ and $\lambda = 1.21$ respectively. (b,e) momentum-space correlations (Eq.~\ref{eq:Ok}) and (c,f) real-space correlations (Eq.~\ref{eq:RealSpCorr}) for $\lambda = 1.19$ S2 phase, and $\lambda = 1.21$ S1 phase (top to bottom panel respectively). All Results are obtained using $64$ sites DMRG simulation. 
    }
\vspace{-0.2cm}
\label{fig:s1}
\end{figure*}

Therefore the Heisenberg Hamiltonian can be written in terms of fermionic operators with some constants that facilitate the representation while doesn't change the symmetry. Some straightforward algebra leads to:
\begin{equation}
	\begin{split}
		H &= 2\sum_{\<ij\>}\textbf{S}_i \cdot \textbf{S}_j \\
		  &= 2 S_i^z S_j^z + (S^+_i S^-_j + S^-_i S^+_j)  \\
		  &= -\sum_{\<ij\>}\Bigl(\sum_{\sigma\sigma^{'}} c_{i\sigma}^\dagger c_{j\sigma} c_{j\sigma '}^\dagger c_{i\sigma '}-\frac{1}{2}n_i n_j + n_i\Bigr)\\
		  &\equiv-\sum_{\<ij\>}\sum_{\sigma\sigma^{'}} c_{i\sigma}^\dagger c_{j\sigma} c_{j\sigma '}^\dagger c_{i\sigma '}
	\end{split}
\end{equation}
which is apparently of SU(2) symmetry.

For spin-1 sites, define spin-1 operator $\textbf{S}_i = \psi_i^\dagger \vec{I} \psi_i$, where $\vec{I}$ is spin matrix in spin-1 Hilbert space:
\begin{equation}
    I^z = \begin{bmatrix}
        1 \;&  0 & 0 \\
        0 \;&  0 & 0 \\
        0 \;&  0 & -1
    \end{bmatrix}, \;
    I^{+} = (I^-)^\dagger = \begin{bmatrix}
            0 \;&  \sqrt{2} & 0 \\
            0 \;&  0 & \sqrt{2} \\
            0 \;&  0 & 0
    \end{bmatrix}
\end{equation}
and $\psi = (d_{i,1},d_{i,0},d_{i,-1})^T$, with $d_{i,m} (d_{i,m}^\dagger)$ being fermion annihilation (creation) operator of spin-m at site $i$. With the constraint that one fermion per site, after some lengthy algebra the ULS Hamiltonian can be written as follows:
\begin{equation}
    \begin{split}
        H_{ULS} &= \sum_{\<ij\>}\Bigl( \textbf{S}_i \cdot \textbf{S}_j + (\textbf{S}_i \cdot \textbf{S}_j)^2 -2 n_i n_j \Bigr)\\
        &= -\sum_{\<ij\>}\Bigl(\sum_{m,m^{'}}(d_{i,m}^\dagger d_{j,m} d_{j,m'}^\dagger d_{j,m'}) - n_in_j + 3 n_i\Bigr)\\
        &\equiv-\sum_{\<ij\>}\sum_{m,m^{'}}d_{i,m}^\dagger d_{j,m} d_{j,m'}^\dagger d_{j,m'}
    \end{split}
\end{equation}
where in the last equivalence we leave out the constant terms $n_i n_j$ and $n_i$. The last expression can be compactly written as $\sum_{\<ij\>} (\psi_i^\dagger \psi_j)(\psi_j^\dagger \psi_i)$, by which one can readily notice the SU(3) symmetry.

\section{Additional results}
In the main text Figure.2 we showed the change of the decay of correlation functions in S2, the phase with a double-peak structure factor, and S1, the novel gapless phase. Here we present the correlation functions near the S1-SOS phase transition. Within S1 phase ($0.34< \lambda < 1.20$), increasing the spin orbital coupling $\lambda$ to S1-SOS transition point blurs the bifurcation structure of $L(k),\;J(k)$ at $k = \pi$ in momentum space correlation. As shown in Figure.S3(b,e), at the end of the S1 phase the correlation in momentum space, though exhibit a small peak at $\pi$ that is indicative of a weak anti-ferromagnetic correlation, is close to the polarized phase in which such structure disappears. Similarly, the alignment of S1 in real space, shown in Figure.3S(c) also exhibits a very small anti-ferromagnetic fluctuation in $L$ and $J$ sectors. This weak anti-ferromagnetic behavior vanishes as one keeps increasing $\lambda$ and reach the SOS phase.

\section{Reproducing data using Itensor}
The full open source code, sample inputs, and corresponding
computational details can be found at \url{https://github.com/fengshi96/hd4-dmrg}

\end{document}